\documentclass[12pt,letter]{article}
\pdfoutput=1
\usepackage{graphicx, epsfig, color,cite}
\usepackage{amsmath}
\usepackage{amssymb}
\usepackage{subfigure}
\def\eslt{\not\!\!\!{E_T}}
\def\to{\rightarrow}

\def\bi{\begin{itemize}}
\def\ei{\end{itemize}}

\def\tu{\tilde u}

\def\tb{\tilde b}

\def\tst{\tilde t}

\def\tg{\tilde g}

\def\tell{\tilde\ell}
\def\tq{\tilde q}
\def\tw{\widetilde W}
\def\tz{\widetilde Z}

\def\alt{\lesssim}
\def\agt{\gtrsim}
\def\be{\begin{equation}}  
\def\ee{\end{equation}}  
\def\bea{\begin{eqnarray}}  
\def\eea{\end{eqnarray}}

\begin{document}
\begin{titlepage}
\begin{flushright}
OU-HEP-171104
\end{flushright}

\vspace{0.5cm}
\begin{center}
{\Large \bf Higgs and superparticle mass predictions \\
from the landscape
}\\ 
\vspace{1.2cm} \renewcommand{\thefootnote}{\fnsymbol{footnote}}
{\large Howard Baer$^1$\footnote[1]{Email: baer@ou.edu },
Vernon Barger$^2$\footnote[2]{Email: barger@pheno.wisc.edu},
Hasan Serce$^3$\footnote[3]{Email: serce@ou.edu}
and 
Kuver Sinha$^1$\footnote[4]{Email: kuver.sinha@ou.edu}
}\\ 
\vspace{1.2cm} \renewcommand{\thefootnote}{\arabic{footnote}}
{\it 
$^1$Dept. of Physics and Astronomy,
University of Oklahoma, Norman, OK 73019, USA \\[3pt]
}
{\it 
$^2$Dept. of Physics,
University of Wisconsin, Madison, WI 53706 USA \\[3pt]
}
{\it 
$^3$Dept. of Engineering and Physics,
University of Central Oklahoma, Edmond, OK 73034 USA \\[3pt]
}

\end{center}

\vspace{0.5cm}
\begin{abstract}
\noindent

Predictions for the scale of SUSY breaking from the string landscape  go back at least a decade to the work of Denef and Douglas on the statistics of flux vacua. 
The assumption that an assortment of SUSY breaking $F$ and $D$ terms are present in the hidden sector, and their values are uniformly distributed in the landscape 
of $D=4$, $N=1$ effective supergravity models, leads to the expectation that the landscape pulls towards large values of soft terms favored by a power law behavior 
$P(m_{soft}) \sim m^n_{soft}$. On the other hand, similar to Weinberg's prediction of the cosmological constant, one can assume an anthropic selection of weak scales 
not too far from the measured value characterized by $m_{W,Z,h} \sim 100$ GeV. Working within a fertile patch of gravity-mediated low energy effective theories where 
the superpotential $\mu$ term is $\ll m_{3/2}$, as occurs in models such as radiative breaking of Peccei-Quinn symmetry, this biases statistical distributions on the 
landscape by a cutoff on the parameter $\Delta_{\rm EW}$, which measures fine-tuning in the $m_Z$-$\mu$ mass relation. The combined effect of statistical and anthropic 
pulls turns out to favor low energy phenomenology that is more or less agnostic to UV physics. While a uniform selection $n=0$ of soft terms produces too low a 
value for $m_h$, taking $n=1$ and $2$ produce most probabilistically $m_h\sim 125$ GeV for negative trilinear terms. For $n\ge 1$, there is a pull towards split generations 
with $m_{\tq,\tell}(1,2)\sim 10-30$ TeV whilst $m_{\tst_1}\sim 1-2$ TeV. The most probable gluino mass comes in at $\sim 3-4$ TeV--apparently beyond the reach of HL-LHC 
(although the required quasi-degenerate higgsinos should still be within reach). We comment on consequences for SUSY collider and dark matter searches. 


\end{abstract}
\end{titlepage}

\section{Introduction}
\label{sec:intro}

One of the great mysteries of fundamental physics is the origin of the vastly 
different energy scales which appear in nature.
Paramount among these is the cosmological constant problem: why is the measured
value of $\Lambda\simeq 10^{-47}$ GeV$^4$ so much smaller than the (reduced) 
Planck scale $M_P^4\simeq 3.3\times 10^{73}$ GeV$^4$? Weinberg proposed an 
anthropic explanation\cite{Weinberg:1987dv}: 
in a vast set of possible universes each with different
(uniformly distributed) possibilities for $\Lambda$, if $\Lambda$ were too
much larger than its measured value, then the universe would expand too rapidly
for galaxies to condense, and the latter constraint seems necessary for the 
appearance of life as we know it. 
Using such reasoning, Weinberg was able to 
predict the value of $\Lambda$ to within a factor of a few of its 
measured value at a time when many physicists expected its value to be zero.
The expectation of a vast set of possible universes (the multiverse) 
found strong support in string theory where stabilization of moduli via 
flux compactifications\cite{Bousso:2000xa,Kachru:2003aw}
led to the emergence of the string theory landscape\cite{Susskind:2003kw}.

Perhaps as intriguing as the cosmological constant problem is the
presence of the gauge hierarchy enigma: why is the weak scale
as typified by $m_{W,Z,h}\sim 100$ GeV so much smaller than the
scale of grand unification $m_{GUT}\simeq 2\times 10^{16}$ GeV when it is well
known that fundamental scalar masses are intrinsically unstable under
quantum corrections\cite{Susskind:1978ms}?
In this case, the expansion of the set of spacetime symmetries in the 
Standard Model (SM) to include supersymmetry (SUSY) results in cancellation of 
quadratic divergences to all orders\cite{Witten:1981nf}. 
The remaining log divergences are relatively mild and at least 
allow for a stable value of the weak scale with the prospect of no funetuning. 
And indeed from this point of view it is possible to view the 
presence of spacetime SUSY with weak scale soft breaking 
as a necessary feature in an anthropic vacuum.

An expansion of the SM to the Minimal Supersymmetric Standard Model (MSSM) 
is actually supported by three disparate data sets: 
1. the measured values of the gauge couplings are exactly what is needed 
for grand unification at a scale $m_{GUT}\simeq 2\times 10^{16}$ GeV, 
2. the measured value of the top quark mass falls in the range needed 
to radiatively break electroweak symmetry in the MSSM and 
3. the measured value of the Higgs boson mass $m_h\simeq 125$ GeV falls
squarely within the predicted narrow MSSM window where $m_h\alt 135$ GeV 
is required\cite{mhiggs}. 
In spite of these successes, so far no signal for
superparticles has yet emerged from dedicated searches by LHC experiments
using  $\sim 100$ fb$^{-1}$ of data from $pp$ collisions at $\sqrt{s}=13$ TeV.
The lack of superpartners at LHC has called into question whether
weak scale SUSY is indeed nature's solution to the naturalness puzzle,
and whether the emergence of a {\it Little Hierarchy} between the
weak scale and the superpartner scale is indicative of the collapse of the SUSY
paradigm\cite{Craig:2013cxa}.

Early calculations of upper bounds on SUSY particles seemed to require 
charginos with mass $m_{\tw_1}\alt 100$ GeV and gluinos with 
$m_{\tg}\alt 350$ GeV\cite{BG}. 
Recently, these calculations have been challenged\cite{comp3,seige} in that they compute using a 
log derivative measure\cite{eenz,BG} 
$\Delta_{BG} \equiv max_i |\frac{\partial\log m_Z^2}{\partial\log p_i}|$ 
in terms of multiple soft terms $p_i$ (assumed independent) whereas in
more fundamental theories the soft terms are all {\it dependent} in that they are 
computable in terms of more fundamental parameters 
(such as the gravitino mass $m_{3/2}$ in gravity-mediated SUSY breaking).
By combining the dependent soft terms, then large cancellations can occur
leading to much less fine-tuning. Other evaluations of fine-tuning
required not-too-large logarithmic corrections to the Higgs mass squared, thus
seemingly requiring three third generation squarks with mass bounded by
500 GeV\cite{dmhiggs}. 
These calculations ignore various
dependent contributions to the renormalization group equation (RGE) of the
up-Higgs soft term $m_{H_u}^2$ which allow for 
{\it radiatively-driven naturalness} wherein large seemingly unnatural 
high scale soft terms such as $m_{H_u}^2$ can be driven by radiative 
corrections to natural values $\sim m_Z^2$ at the weak scale.

An improved naturalness measure $\Delta_{\rm EW}$ has been proposed\cite{ltr,rns} which just 
requires that {\it weak scale} contributions to $m_Z^2$ should be 
comparable to or less than $m_Z^2$. From the minimization conditions for the MSSM Higgs potential\cite{WSS} one finds
\be 
\frac{m_Z^2}{2} = \frac{m_{H_d}^2 + \Sigma_d^d -
(m_{H_u}^2+\Sigma_u^u)\tan^2\beta}{\tan^2\beta -1} -\mu^2 \simeq 
-m_{H_u}^2-\Sigma_u^u-\mu^2 .
\label{eq:mzs}
\ee 
The radiative corrections $\Sigma_u^u$ and $\Sigma_d^d$ include contributions 
from various particles and sparticles with sizeable Yukawa and/or gauge
couplings to the Higgs sector.
Usually the most important of these are
\be
\Sigma_u^u (\tst_{1,2})= \frac{3}{16\pi^2}F(m_{\tst_{1,2}}^2)
\left[ f_t^2-g_Z^2\mp \frac{f_t^2 A_t^2-8g_Z^2
(\frac{1}{4}-\frac{2}{3}x_W)\Delta_t}{m_{\tst_2}^2-m_{\tst_1}^2}\right]
\label{eq:Sigmat1t2}
\ee
where $f_t$ is the top-quark Yukawa coupling, 
$\Delta_t=(m_{\tst_L}^2-m_{\tst_R}^2)/2+M_Z^2\cos 2\beta (\frac{1}{4}-\frac{2}{3}x_W)$, $x_W\equiv\sin^2\theta_W$, 
$F(m^2)= m^2\left(\log\frac{m^2}{Q^2}-1\right)$ and
the optimized scale choice for evaluation of these corrections is
$Q^2=m_{\tst_1}m_{\tst_2}$.
In the denominator of Eq.~\ref{eq:Sigmat1t2},
the tree level expressions of $m_{\tst_{1,2}}^2$ should be used.
Expressions for the remaining $\Sigma_u^u$ and $\Sigma_d^d$ terms 
are given in the Appendix of Ref. \cite{rns}.

The naturalness measure $\Delta_{\rm EW}$ 
compares the largest contribution on the right-hand-side of Eq. \ref{eq:mzs} 
to the value of $m_Z^2/2$. 
If they are comparable ($\Delta_{\rm EW}\alt 30$), 
then no unnatural fine-tunings are required to generate $m_Z=91.2$ GeV. 
The main requirement for low fine-tuning is then that
\bi
\item $|\mu |\sim 100-300$ GeV\cite{Chan:1997bi,Barbieri:2009ew,hgsno} (the lighter the more natural with $\mu \agt 100$ GeV to accommodate LEP2 limits 
from chargino pair production searches).
\item  Also, $m_{H_u}^2$ is driven radiatively to small 
$(\sim -(100-300)^2$ GeV$^2$, and not large, negative values~\cite{ltr,rns}. 
\item The top squark contributions to the radiative corrections $\Sigma_u^u(\tst_{1,2})$ 
are minimized for TeV-scale highly mixed top squarks\cite{ltr}. 
This latter condition  also lifts the Higgs mass  to $m_h\sim 125$ GeV.
\item First and second generation squark and slepton masses may range as 
high as 10-30 TeV with little cost to naturalness\cite{rns,upper}.
Such a high mass range offers a decoupling solution to 
the SUSY flavor, CP and gravitino problems\cite{dine_etc}.
\ei

The question then arises: why should the soft SUSY breaking terms and the 
superpotential $\mu$ term adopt the specific range of values needed to satisfy the naturalness condition?

In the case of the $\mu$ term, it has been commonly assumed that $\mu$ takes a value
comparable to the SUSY breaking scale as suggested in the Giudice-Masiero mechanism\cite{GM}.
If that were so-- and  with soft terms now required to lie in the multi-TeV regime by LHC 
constraints-- then one would have to accept a multi-TeV value of $\mu$ and the MSSM
would necessarily be fine-tuned with $|\mu|\gg m(W,Z,h)$. 
However, in the Kim-Nilles (KN) $\mu$ term solution\cite{KN}, 
which is a supersymmetrized version of the DFSZ axion model\cite{dfsz}, 
the expectation can be very different. In KN, the Higgs superfields carry a common
PQ charge so that the $\mu$ term is initially forbidden by PQ symmetry. 
Upon spontaneous PQ symmetry breaking, an axion is generated to solve the strong CP
problem but also a $\mu$ parameter is generated with value $\mu\sim \lambda_{\mu}v_{PQ}^2/m_P$\footnote{
In addition, an intermediate scale Majorana neutrino mass $m_N$ is also generated.}.
This may be compared to the SUSY breaking scale $m_{soft}\sim m_{hidden}^2/m_P$ 
where $m_{hidden}$ is some intermediate mass scale associated with the hidden sector.
Then $\mu\ll m_{soft}$ is just a consequence of $v_{PQ}< m_{hidden}$. Indeed, in 
models of radiative PQ breaking\cite{msy}, 
wherein PQ breaking is derived as a consequence of SUSY breaking, then typically
$\mu\ll m_{soft}$ is expected\cite{radpq}. 
We note here that $\mu\ll m_{soft}$ arises in other well-motivated cases, 
such as certain classes of string models with flux compactifications\cite{quevedo}.

Regarding natural values for the soft SUSY breaking terms, one possibility is that, with the
right correlations amongst soft terms and a small superpotential $\mu$ term $\sim 100-300$ GeV, 
then a generalized focus point mechanism\cite{fp} can exist such that $m_{H_u}^2$ runs to small negative values
at the weak scale roughly independently of its high scale value\cite{gfp}. 
Another possibility arises from the string theory landscape. 
If-- within a ``fertile patch'' of the landscape of string theory vacua (such that the low energy effective theory
is the MSSM or related variants)-- there is 
\begin{enumerate}
\item a statistical selection towards large soft terms\cite{denefdouglas,doug1,dine} and 
\item an anthropic selection towards a weak scale value $m(W,Z,h)$ not too far from 
$\sim 100$ GeV\cite{Agrawal:1997gf} and 
\item a mechanism such as radiative PQ breaking which generates 
$\mu\sim m(weak)$ rather than $\mu\sim m_{soft}$, 
\end{enumerate}
then the soft terms are pulled towards those values which generate natural SUSY in accord with 
Eq. \ref{eq:mzs} and a light Higgs mass $m_h\simeq 125$ GeV\cite{lscape}.\footnote{
Condition \#1, as argued by Denef and Douglas, seems generic in string theory. 
Condition \#2 may\cite{Agrawal:1997gf} or may not be generic in string theory vacua.
Condition \#3 emerges from the assumed solution to the SUSY $\mu$ problem.
Weak scale naturalness prefers $\mu\sim m_{W,Z,h}$ while LHC results prefer the SUSY 
breaking scale $m_{3/2}$ in the multi-TeV regime.
Since the MSSM $\mu$ term is supersymmetric and not SUSY breaking, a solution to the SUSY
$\mu$ problem, such as Kim-Nilles\cite{KN} where $\mu$ can be $\ll m_{3/2}$ (while solving the
strong CP problem and generating intermediate scale right-hand Majorana neutrino masses) 
seeems preferred to us over other mechanisms which generate $\mu\sim m_{3/2}$. 
For further discussion, see {\it e.g.} Ref.~\cite{radpq}.} 
The combined draw--  
1. towards large soft terms and 2. towards an anthropic weak scale-- 
pulls the high scale value of 
$m_{H_u}^2$ to such large values that electroweak symmetry is ``barely broken''\cite{GR}. 
This is the same as the naturalness condition
that $m_{H_u}^2$ be driven to small negative values at the weak scale.

While Ref.~\cite{lscape} provided a qualitative picture for understanding why the soft terms adopt values
required for naturalness, in the present work we attempt to place this approach on a more quantitative footing.
In Sec. \ref{sec:string}, we review some ideas mainly originating from 
Douglas and Denef regarding the draw of the string theory landscape 
towards large soft SUSY breaking terms as described by a power law selection $f_{SUSY}(m_{soft})\sim
m_{soft}^n$. A mild pull towards large soft SUSY breaking terms comes from values of 
$n\sim 1$ or 2 which arises from rather simple hidden sectors where SUSY breaking arises 
from just one or two fields gaining a SUSY breaking vev. In contrast, larger values of 
$n\ge 3$ emerge from more complicated hidden sectors where several or more fields gain 
comparable SUSY breaking vevs and thus exert a stronger pull towards large 
values of soft breaking terms. 
We combine this with an anthropic draw towards the measured value of the weak scale. 
The combination of both allows us to calculate probability distributions for
expected Higgs boson and superparticle masses.
In Sec. \ref{sec:num}, we implement this methodology with its power law selection for 
large soft terms which are then passed on to the SUSY spectrum generator contained in Isajet 7.87\cite{isajet}. 
By assuming a $\mu$ parameter not too far from $m_{weak}$, then we are able to {\it invert} 
the normal useage of Eq. \ref{eq:mzs} to calculate the value of $m_Z$ which is in general
{\it not} equal to its measured value. 
If $m_Z$ is too large, then also the weak scale is too large, thus
suppressing rates for weak interactions and increasing particle masses which arise from 
electroweak symmetry breaking. 
Requiring that the weak scale not deviate by more than a factor of 
a few from its measured value (in accord with calculations from Agrawal 
{\it et al.}\cite{Agrawal:1997gf}), 
then we are able to present our results as probability distributions versus various observable masses.
Some confidence in this approach is gained in that the 
probability distribution for the light Higgs mass peaks rather sharply at $m_h\sim 125$ GeV. 
It is intriguing that this already occurs for the simplest case of SUSY breaking which is 
dominated by a single $F$-term field which yields $n=1$.
We then also find $m_{\tg}\sim 3-4$ TeV and $m_{\tst_1}\sim 1-2$ TeV.
First/second generation scalar masses are pulled into the 10-30 TeV range leading to 
an amelioration of the SUSY flavor and CP problems. 
Higher values of $n\ge 3$ tend to pull the soft terms to such large values that one is placed into
charge or color breaking (CCB) electroweak vacua or else vacua where electroweak symmetry 
doesn't even break.
In Sec. \ref{sec:col_dm} we discuss some inplications of our results 
for collider searches for SUSY and for dark matter searches for WIMPs and axions.
In Sec. \ref{sec:cosmo} we discuss some aspects of the cosmological moduli problem and in Sec. \ref{sec:conclude} we present a summary and conclusions.

\section{String vacuum statistics and the SUSY breaking scale}
\label{sec:string}

In this Section, we assume a vast ensemble of string vacua states which give rise to a $D=4$, $N=1$
supergravity effective field theory at high energies. Furthermore, 
the theory consists of a visible sector containing the MSSM along with a perhaps large assortment of
fields that comprise the hidden sector. 
The scalar potential is given by the usual supergravity form\cite{Nilles:1983ge}
\bea
V&=&e^{K/m_P^2} \left( g^{i \overline{j}} D_{i}W D_{\overline{j}}W^{*} \, - \, 
\frac{3}{m_P^2} |W|^2 \right) \, + \, \frac{1}{2}\sum_\alpha D^2_\alpha \,\, \\
&=&e^{K/m_P^2} \left( \sum_i|F_i|^2-3\frac{|W|^2}{m_P^2}\right)+\frac{1}{2}\sum_\alpha D_\alpha^2
\eea
where $W$ is the holomorphic superpotential, $K$ is the real K\"ahler potential\footnote{
Not to be confused with the (dimensionless) K\"ahler function $G=K/m_P^2+\log |W/m_P^3|^2$.} 
and $F_i=D_iW=DW/D\phi^i\equiv\partial W/\partial\phi^i+(1/m_P^2)(\partial K /\partial\phi^i)W$
are the $F$-terms and $D_\alpha\sim \sum\phi^\dagger gt_{\alpha}\phi$ are the $D$-terms
and the $\phi^i$ are chiral superfields.
Supergravity is assumed to be broken spontaneously via the super-Higgs mechanism either via
$F$-type breaking or $D$-type breaking or in general a combination of both leading to a gravitino mass
$m_{3/2}=e^{K/2m_P^2}|W|/m_P^2$. 
The (metastable) minima of the scalar potential can be found by requiring
$\partial V/\partial\phi^i=0$ with $\partial^2V/\partial\phi^i\partial\phi^j>0$ to ensure a local
minimum.
The cosmological constant is given by
\be
\Lambda_{cc} \,\, = \,\, m_{hidden}^4\, - \, 3 e^{K/m_P^2}|W|^2/m_P^2 \,\, 
\ee
where $m_{hidden}^4=\sum_i |F_i|^2 \, +  \, \frac{1}{2}\sum_\alpha D^2_\alpha $ is a mass scale associated
with the hidden sector (and usually in SUGRA-mediated models it is assumed $m_{hidden}\sim 10^{12}$ 
GeV such that the gravitino gets a mass $m_{3/2}\sim m_{hidden}^2/m_P$).

A key observation of Susskind\cite{suss} and Denef and Douglas\cite{denefdouglas,doug1} (DD) 
was that $W$ at the minima is distributed uniformly as a complex variable, and the distribution of 
$e^{K/m_P^2}|W|^2/m_P^2$ 
is not correlated with the distributions of $F_i$ and $D_\alpha$. 
Setting the cosmological constant to nearly zero, then, has no effect on the distribution of 
supersymmetry breaking scales. 
Physically, this can be understood by the fact that the superpotential receives contributions 
from many sectors of the theory, supersymmetric as well as non-supersymmetric. 

\begin{figure}[tbp]
\begin{center}
\includegraphics[height=0.3\textheight]{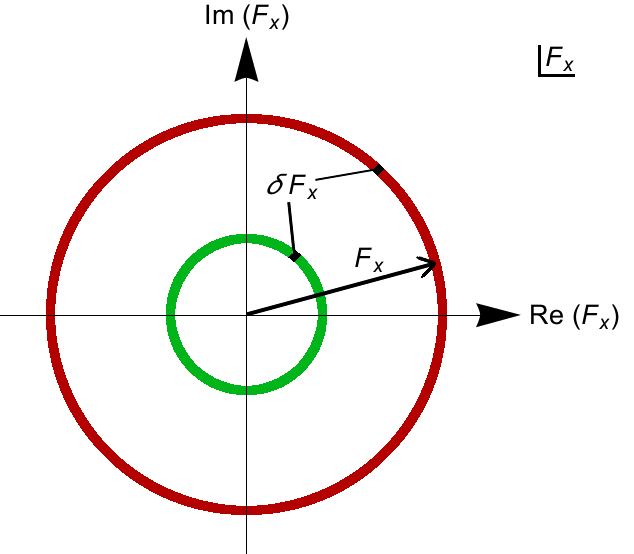}
\caption{
Annuli of the complex $F_X$ plane giving rise to linearly increasing selection of soft SUSY breaking terms.
\label{fig:f_x}}
\end{center}
\end{figure}

Next, we would like to estimate the number of flux vacua containing spontaneously broken
supergravity with a SUSY breaking scale $m_{hidden}^2$, $dN_{vac}[m_{hidden}^2,m_{weak},\Lambda ]/dm_{hidden}^2$.
According to DD\cite{doug1,doug2,DK,Kumar:2006tn}, this distribution is likely to be the product of three factors:
$f_{SUSY}(m_{hidden}^2)$, $f_{EWFT}$ and $f_{cc}$. 
\be
dN_{vac}[m_{hidden}^2,m_{weak},\Lambda ]=f_{SUSY}(m_{hidden}^2)\cdot f_{EWFT}\cdot f_{cc}\cdot dm_{hidden}^2 
\label{eq:dNvac}
\ee
which contain $\Lambda\sim 0$ but with $m_{weak}\simeq m_{W,Z,h}\sim 100$ GeV.
The cosmological fine-tuning penalty is $f_{cc}\sim \Lambda/m^4$ where the above discussion
leads to $m^4\sim m_{string}^4$ rather than $m^4\sim m_{hidden}^4$, rendering this term inconsequential for
determining the number of vacua with a given SUSY breaking scale.
Another key observation from examining flux vacua in IIB string theory 
is that the SUSY breaking $F_i$ and $D_\alpha$ terms
are likely to be uniformly distributed-- in the former case as complex numbers while in the latter case
as real numbers.  
In this case, one then obtains the following distribution of supersymmetry breaking scales
\be 
\label{brokensusydistri}
f_{SUSY}(m_{hidden}^2) \, \sim \, (m^2_{hidden})^{2n_F+n_D - 1}
\ee
where $n_F$ is the number of $F$-breaking fields and $n_D$ is the number of $D$-breaking fields in
the hidden sector\cite{doug1}.
The case of $n_F = 1$ is displayed in Figure~\ref{fig:f_x}. 
We label the visible sector soft term mass scale as $m_{soft}$ where in SUGRA breaking models 
we typically have $m_{soft}\sim m_{hidden}^2/m_P\sim m_{3/2}$. 
Thus, the case of $n_F=1$ $n_D=0$ would give a {\it linearly increasing} probability distribution 
for generic soft breaking terms simply because the area of annuli within the complex plane increases
linearly. We will denote the collective exponent in Eq. \ref{brokensusydistri} as 
$n\equiv 2n_F+n_D-1$ so that the case $n_F=1$, $n_D=0$ leads to $n=1$ with 
$f_{SUSY}(m_{soft})\sim m_{soft}^1$. 
The case $n_F=0$ with $n_D=1$ would lead to a uniform distribution in soft terms 
$f_{SUSY}(m_{soft})\sim m_{soft}^0$. For the more general case with an assortment of $F$ and $D$ terms 
contributing comparably to SUSY breaking, then high scale SUSY breaking models 
would be increasingly favored.\footnote{The authors of Ref. \cite{kai} argue that that low scale
SUSY breaking is preferred by the cosmological constant\cite{Banks:2003es} 
but then possible formation of 
cosmological domain walls via $R$-symmetry breaking provides a lower bound on the 
scale of SUSY breaking and hence upon $m_{3/2}$.}
\begin{table}[htb]
\begin{center}
\begin{small}
\begin{tabular}{|ccc|}
\hline
$n_F$ & $n_D$ & $n$ \cr
\hline
0 & 1 & 0 \\
1 & 0 & 1 \\
0 & 2 & 1 \\
1 & 1 & 2 \\
0 & 3 & 2 \\
2 & 0 & 3 \\
2 & 1 & 4 \\
\hline 
\end{tabular}
\end{small}
\smallskip
\caption{Some choices of $n_f$ and $n_D$ leading to different $n$ values.}
\label{tab:n}
\end{center}
\end{table}

The third factor in the SUSY breaking distribution $f_{EWFT}(m_{soft})$ arises from anthropics and places
a penalty on the calculated value of the weak scale deviating too much from its measured 
value $m_{weak}\sim 100$ GeV. Following \cite{splitsusy}, DD advocated the form\cite{doug2}
\be
f_{EWFT}\sim m_{weak}^2/m_{soft}^2
\label{eq:fDD}
\ee
so that the more the soft terms increase beyond the weak scale, the greater is the penalty. 
This factor must be interpreted with some care. 
At first glance, one would expect that the larger the value of $m_{soft}$ becomes, 
then the larger is the calculated value of the weak scale.
However, this does not hold true for a variety of cases.
\begin{itemize}
\item In one case, as trilinear soft terms increase, then the visible sector scalar potential
develops charge and/or color breaking (CCB) minima (see Fig. 1 of \cite{lscape}), 
leading to a universe not as we know it, 
and likely not conducive to observers. Another possibility is that as soft terms such as
$m_{H_u}^2$ increase relative to other soft terms, then its value is too large to be driven
radiatively to negative values so that electroweak symmetry doesn't even break. Such string vacua--
even within the context of spontaneously broken SUGRA in the MSSM+hidden sector paradigm-- 
must be vetoed by our selection rules.
\item Even in the case where EW symmetry is properly broken, it is not always the case that
increasing soft terms lead to larger values of the calculated weak scale. One case consists of
the soft term $m_{H_u}^2$: the larger its high scale value becomes, 
then the larger is its cancelling correction from radiative corrections/RG running\cite{savoy}. 
For too small values of $m_{H_u}^2$, then it runs deeply negative at the weak scale leading to
some required fine-tuning by adopting a large value of $\mu$ to compensate and keep $m_Z$ or
$m_h$ at its measured value. But for larger values of $m_{H_u}^2(m_{GUT})$, then $m_{H_u}^2$
runs to small weak scale values, thus barely breaking EW symmetry\cite{GR,lscape}.
For yet higher values of $m_{H_u}^2(m_{GUT})$, then $m_{H_u}^2$ doesn't even run negative at the weak
scale, and EW symmetry remains unbroken.

Another case consists of the trilinear soft term $A_t$. For small values of $A_t$, 
then there is little mixing in the stop sector. Not only is it difficult to raise $m_h$ up to
its measured value\cite{h125}, but the radiative corrections $\Sigma_u^u(\tst_{1,2})$ in Eq. \ref{eq:mzs}
become large, leading to either large fine-tuning, or in the case where
$\mu$ is fixed and $m_{weak}$ floats, to a too large value of $m_{weak}$. 
As the weak scale value of $A_t$ increases, then large cancellations occur 
in both $\Sigma_u^u(\tst_1 )$ and $\Sigma_u^u(\tst_2)$ leading to greater naturalness 
and an increased $m_h\sim 125$ GeV\cite{ltr}.
\end{itemize}

\subsection{$f_{EWFT}$: case {\bf A}}

To ameliorate this situation, we advocate two different replacements of Eq. \ref{eq:fDD}.
\be
case\ {\bf A}:\ \ \ f_{EWFT}\to \Theta (30 -\Delta_{\rm EW}) ,
\label{eq:caseA}
\ee
where $\Theta (x)$ is the usual Heaviside unit step function $\Theta (x)=0\ (1)$ for $x\le 0$ ($x>0$).
In our methodology, we assume $\mu$ is generated to small values not too far from $m_{weak}$
but then we invert the usual useage of Eq. \ref{eq:mzs} to let $m_Z$ float so that 
large values of $\sqrt{|m_{H_u}^2(weak)|}$ or $\Sigma_u^u$ generate large values of the
weak scale $m_{weak}\gg 100$ GeV. The value of $\Delta_{\rm EW}<30$ then corresponds to
calculated anthropic requirements from Agrawal {\it et al.} that the weak scale not deviate 
by more than a factor of several from its measured value\cite{Agrawal:1997gf}. 
In this case, $\Delta_{\rm EW}=30$ corresponds
to a $Z$ mass nearly four times its measured value.

\subsection{$f_{EWFT}$: case {\bf B}}

We also examine
\be
case\ {\bf B}:\ \ \ f_{EWFT}\to \Delta_{\rm EW}^{-1}
\label{eq:caseB}
\ee
which is more closely tied to the DD prescription in that
\be
\Delta_{\rm EW}^{-1}\sim (m_Z^2/2)/max\left[ |m_{H_u}^2(weak)|\ or\ \mu^2\ \ or\ \ 
|\Sigma_u^u (i)|\right] .
\label{eq:caseBp}
\ee
Instead of placing a generic $m_{soft}^2$ in the denominator of Eq.~\ref{eq:caseBp}, 
we place the maximal weak scale contribution to the magnitude of the weak scale.
Rather than placing a sharp cutoff on the calculated magnitude of the weak scale as in Case {\bf A},
case {\bf B} places an increasing rejection penalty the more the calculated value of the weak 
scale strays from its measured value. 
However, calculated values which differ by large factors from the measured weak scale 
are nonetheless sometimes allowed.

\subsection{Some general comments}

The purpose of the present work is to explore several questions that emerge in this framework. 
On the one hand, the pull towards large supersymmetry breaking scales is evident from 
Eq.~\ref{brokensusydistri}, especially for a large number $n_F$ and/or $n_D$ of SUSY breaking fields. 
Already at $n_F = 2$, a distribution $f_{SUSY}\sim m^3_{soft}$ emerges that is heavily biased towards 
high scale supersymmetry breaking. 
This leads to the question of whether one should expect to see any signatures of supersymmetry 
at low energies, since, naively, the soft terms in the infrared (IR or weak scale) should similarly be
 pulled to larger and larger values. 
On the other hand, one could also ask how predictive low-energy phenomenology is for a given 
scale of SUSY breaking $m_{hidden}$. 
A given scale $m_{hidden}$ can accommodate various statistical distributions corresponding to the 
different powers $n_F$ or $n_D$ in Eq.~\ref{brokensusydistri}. 
Naively, superpartner masses in the IR should show a corresponding statistical distribution, 
raising the question of predictive power. 
For the case of the Higgs mass, which receives corrections from the supersymmetric spectrum, 
the question becomes even more critical - can one argue for a natural value preferred from the landscape?

While the statistical distribution $f_{SUSY}$ clearly pulls $m_{hidden}$ 
(and hence soft masses in the IR) to large values, the imposition of additional constraints 
can balance this effect. 
The most important constraint may be anthropic in nature: it is that the calculated value of the weak scale 
not deviate from its measured value by more than a factor of several. 
Calculations by Agrawal {\it et al.} maintain that anthropically the weak scale should not deviate 
by more than a factor 5 from its measured value: we will adopt a slightly more conservative bound
\be
m_{weak} \, \sim \, m_{W,Z,h} \, \alt 350 {\rm GeV} 
\ee
corresponding to $\Delta_{\rm EW}\alt 30$.
This rests on the observation that rates of nuclear fusion processes and beta decays scale as 
$\sim 1/m^4_{weak}$, and a large value of $m_{weak}$ would severely alter the production of 
heavy elements during Big Bang Nucleosynthesis and in stars. 
A higher weak scale, with all other constants remaining the same, would also result in heavier particles
which receive mass from EWSB. Susskind suggests that the increased masses would speed up 
numerous astrophysical processes\cite{suss} (for more details on astrophysical constraints on a
too-large weak scale, see Ref's \cite{Agrawal:1997gf,kribs}). 
A caveat that should be kept in mind is that this conclusion is true if the weak scale is the 
only parameter that is varied; for example, if one is also allowed to sample other technically 
natural parameters of the Standard Model, perfectly habitable vacua where the Higgs mass 
resides near the Planck scale may be obtained (the so-called ``Weakless Universe" models \cite{kribs}). 
Nevertheless, small fermion masses are more likely to be obtained in a chiral rather than vector theory. 

In the context of supersymmetry, the requirement of an anthropic weak scale can be expressed 
as a concrete requirement on superpartner masses, namely, that the naturalness parameter 
should satisfy $\Delta_{\rm EW} < 30$. 
A natural Universe where supersymmetry resolves the hierarchy problem, then, 
would be one in which $\Delta_{\rm EW} \leq \mathcal{O}(10)$, not only in our vacuum, 
but also in vacua like ours. 
This would ensure that all terms in Eq.~\ref{eq:mzs} are not too far above the measured value of the weak scale.
The distribution of vacua in Eq.~\ref{eq:dNvac} can then be usefully written as
\be \label{fullmeasure}
dN_{vac} \, \sim \,  \Theta (30 - \Delta_{\rm EW}) \times (m^2_{hidden})^n d(m^2_{hidden})  \,\,
\ee
where $n=2n_F+n_D - 1$.
This is a mathematical statement of the strongest sense in which supersymmetry can be taken as a 
solution to the gauge hierarchy problem while not also generating a Little Hierarchy where
$\Delta_{\rm EW}\gg 30$. 

We note that our philosophy with regard to the landscape is similar to the one pursued by 
Douglas \cite{doug2}, with the difference being what we consider to be the correct measure of naturalness. 
In Douglas's 2012 paper \cite{doug2}, the measure adopted was simply 
$f_{EWFT}=m_{weak}^2/m_{soft}^2$. 
Naturalness quantified in this manner is clearly in tension with the findings of the LHC so far, 
since mass limits on gluinos (top squarks) exceed 2 TeV (1 TeV). 

Adopting, instead, the more robust measure $\Delta_{\rm EW}$, we see that the expected low-energy mass spectrum 
is the one described in the Introduction. 
The question then arises: how robust is the expected natural spectrum against different 
values of $n$ in Eq.~\ref{fullmeasure}? 
This isn't an entirely trivial question. 
There are two tendencies in Eq.~\ref{fullmeasure} - the first is the pull towards heavier scalars as 
increasing $n$ pulls the distribution towards larger $m_{soft}$. 
In fact, there is no reason to expect that only one field dominates supersymmetry breaking in the hidden sector. 
On the other hand, however, increasing $n$ tends to increase contributions to the 
radiative corrections $\Sigma_u^u$ and $\Sigma_d^d$ on the right hand side of Eq.~\ref{eq:mzs} 
which pulls the calculated value of $m_{weak}$ beyond its measured value. 
The step function in Eq.~\ref{fullmeasure} then rejects these vacua through the anthropic weak scale. 
In fact, it is not only the low $m_{weak}$ requirement that rejects these vacua - 
many of them are unacceptable because they fall into color-breaking minima or do not break 
electroweak symmetry at all. 
It is thus clear that some distribution of soft masses, centered around a presumably natural set of values, 
is expected as one increases $n$. 
We now go on to show that this is indeed the case. 

\section{Numerical results}
\label{sec:num}

A quantitative investigation of these questions will require us to work within a particular 
mediation scheme with suitable boundary conditions at the GUT scale. 
We choose gravity mediation and a selection of soft terms following the NUHM3 
(three-extra-parameter non-universal Higgs) model\cite{nuhm2} 
although our broad conclusions are independent of specific UV boundary conditions for the soft terms.
The NUHM3 model is convenient in that it allows for $\mu$ as an independent input parameter,
and since we require $\mu$ not too far from $m_{W,Z,h}\sim 100$ GeV.
The NUHM3  model is inspired by previous work on 
mini-landscape investigations of heterotic string theory compactified on a $Z_6-II$ 
orbifold\cite{mini}.
In these models, sparticle masses are dictated by the geography of their wavefunctions within the
compactified manifold. These models exhibit {\it localized} grand unification\cite{local} 
wherein the first/second generation matter superfields lie near fixed points (the twisted sector) 
and thus lie in {\bf 16}-dimensional spinor reps of SO(10). 
Meanwhile, third generations fields and Higgs and vector boson multiplets lie more in the 
bulk and thus occur in split multiplets (solving the doublet-triplet splitting problem) 
and receive smaller soft masses\cite{nilles_kane}. 
Such a set-up motivates the NUHM3 model with the following
parameters $m_0(1,2),\ m_0(3),\ m_{1/2},\ A_0,\ \tan\beta,\ m_{H_u},\ m_{H_d} $
where all mass parameters are taken as GUT scale values.
The soft Higgs masses can be traded for weak scale values of $\mu$ and $m_A$.
Thus, the final parameter space is taken as
\be
m_0(1,2),\ m_0(3),\ m_{1/2},\ A_0,\ \tan\beta,\ \mu ,\ m_A\ \ \ ({\rm NUHM3})
\ee
With the gravitino mass $m_{3/2}\sim m_{hidden}^2/m_P$, then we will adopt
\bea  \label{eq:m32}
m_0(1,2) & = &  c_{1,2} \, \times \, m_{3/2} \nonumber \\
m_0(3) & = &  c_3 \, \times \, m_{3/2} \nonumber \\
m_{1/2} & = &  c_{1/2} \, \times \, m_{3/2}\\
A_0 &=  & -c_{A_0} \, \times \, m_{3/2}\nonumber  \\
m_A &=  & c_{A} \, \times \, m_{3/2}\nonumber
\eea
{\it i.e.} each of these mass terms will scan as $m_{soft}^n$.

We scan according to $m_{soft}^n$ over:
\begin{itemize}
\item $m_0(1,2):\ 0.1 - 60$ TeV,
\item $m_0(3):\ 0.1 - 20$ TeV,
\item $m_{1/2}:\ 0.5 - 10$ TeV,
\item $A_0:\ -50 -\ 0$ TeV,
\item $m_A:\ 0.3 - 10$ TeV,
\end{itemize}
with $\mu = 150$ GeV while $\tan\beta:3-60$ scanned uniformly. 
The goal here is to choose upper limits to our scan parameters
which will lie beyond the upper limits imposed by the anthropic selection from $f_{EWFT}$.
Lower limits are motivated by current LHC search limits. 
Our final results will hardly depend on the chosen value of $\mu$ so long as 
$\mu$ is with an factor of a few of $m_{W,Z,h}\sim 100$ GeV.

\subsection{Case {\bf A}:}

While $\mu$ is fixed to be small, nonetheless large values of $\Delta_{\rm EW}$ can still be generated.
This often occurs due to large contributions to $\Delta_{\rm EW}$ from $m_A/\tan\beta$ or large
contributions to $\Sigma_u^u (\tst_{1,2})$. Usually, in such cases the value of $m_{H_u}^2(weak)$ is 
adjusted/fine-tuned to guarantee that $m_Z$ lies at its measured value. Then
$m_{H_u}^2$ is run back up to $Q=m_{GUT}$ to whatever value is consistent with its weak scale value. 
Alternatively, if we do not fine-tune $m_{H_u}^2(weak)$, then the weak scale will attain a value
\be
m(weak)\simeq \sqrt{\Delta_{\rm EW}\cdot m_Z^2/2} .
\ee
The procedure followed in case {\bf A} is to not tune $m_{H_u}^2(weak)$ and then reject
solutions with $\Delta_{\rm EW}>30$ which would generate a weak scale $m_Z\agt 350 $ GeV, 
nearly four times the measured value of the $Z$ mass.
\begin{figure}[t]
  \centering
  {\includegraphics[width=.48\textwidth]{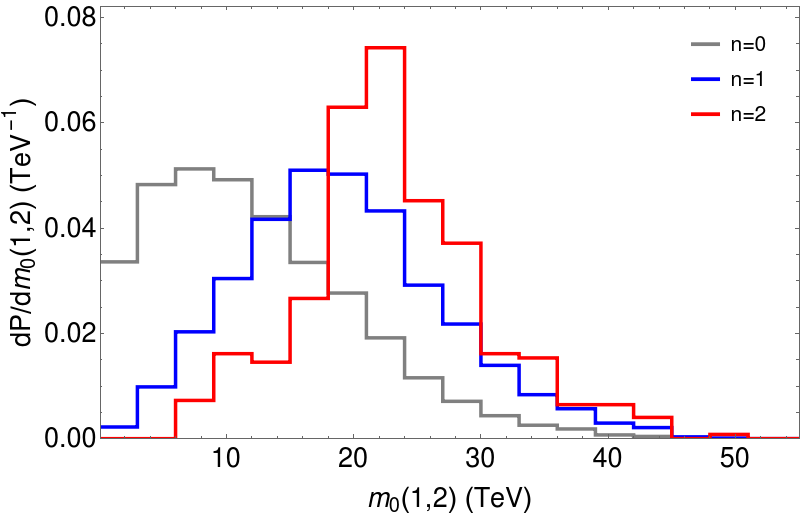}}\quad
  {\includegraphics[width=.48\textwidth]{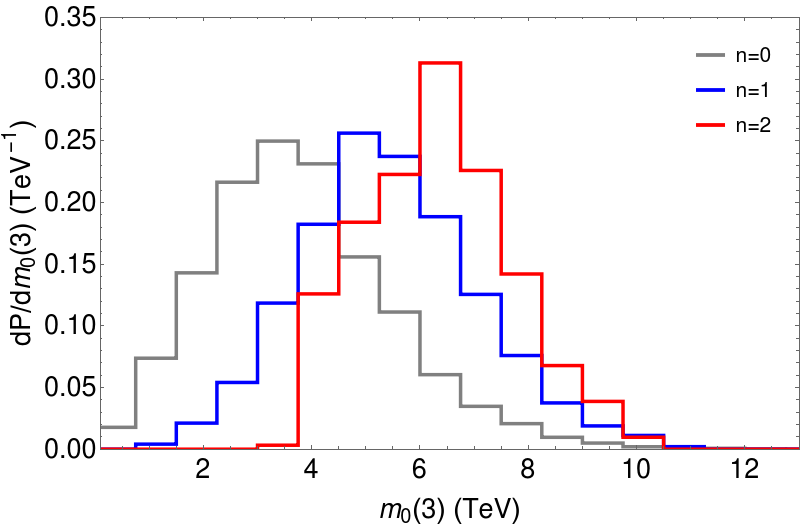}}\\ 
  {\includegraphics[width=.48\textwidth]{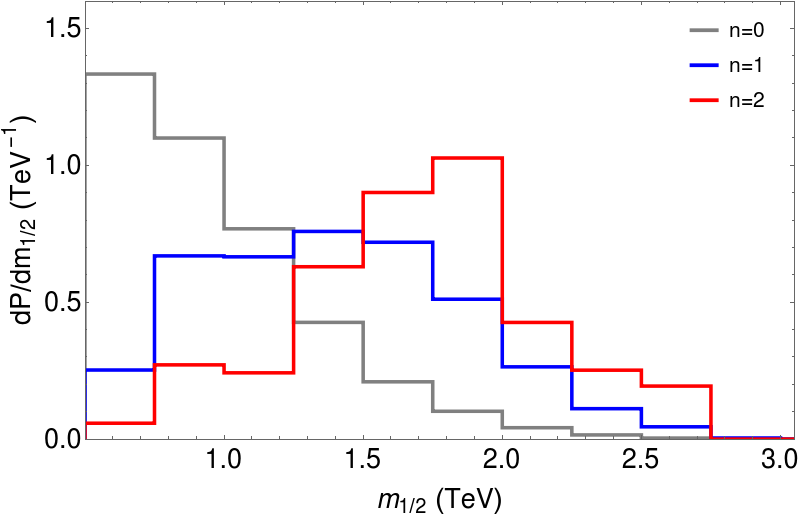}} \quad 
  {\includegraphics[width=.48\textwidth]{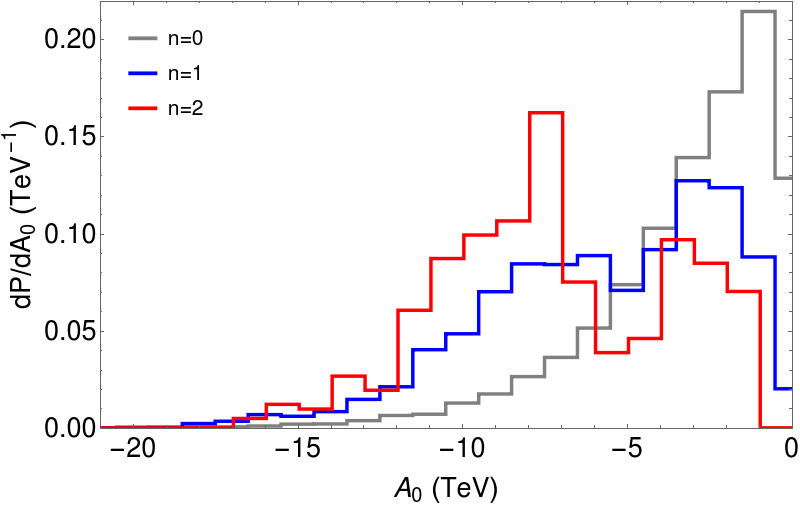}}\\
  \caption{Case\ {\bf A}:\ \ \ $f_{EWFT}\to \Theta (30 -\Delta_{\rm EW})$ - Upper panels: 
Probability distributions in $m_0(1,2)$ (left) and $m_0(3)$  (right). 
Lower panels: Probability distributions in $m_{1/2}$  (left) and $A_0$ (right). 
All distributions are shown following Eq.~\ref{fullmeasure} and in addition rejecting 
non-standard scalar potential minima. 
Results for different values of $n = 2n_F +n_D-1$ are displayed for each plot.  
  }  
  \label{fig:2}
\end{figure}

In Figure~\ref{fig:2}, we plot the probability distributions from our statistical scan over soft terms
versus first/second generation scalar mass $m_0(1,2)$ and third generation soft mass $m_0(3)$ 
in the top panels. 
For the generation 1,2 soft SUSY breaking matter scalar masses, we immediately see from
frame {\it a}) that for the cases
$n=1$ and 2 that the probability distributions peak in the vicinity of $m_0(1,2)\sim 20$ TeV with tails
extending out to 30 TeV. Such large scalar masses occur because of the linear ($n=1$) 
and quadratic $(n=2$) pull on these soft terms with only minimal suppression which sets in at 
$m_0(1,2)\agt 20$ TeV. One avenue for suppression arises from electroweak $D$-term contributions to the 
$\Sigma_{u,d}^{u,d}$ terms which depend on weak isospin and electric charge assignments. 
For nearly degenerate scalars of each generation, these nearly cancel out\cite{maren}. Another avenue for
suppression comes from two loop terms in the MSSM RGEs\cite{mv}: 
if scalar masses enter the multi-TeV range,
then these terms can become large and help drive third generation scalar masses tachyonic leading to
CCB minima in the scalar potential\cite{imh}. 
Both these rather mild suppressions are insufficient to prevent
first/second generation scalar masses from rising to the 20-30 TeV range. Such heavy scalars go a long
way to suppressing possible FCNC and CP violating SUSY processes\cite{dine_etc}.
For the $n=0$ case, $dP/dm_0(1,2)$ peaks around 5-10 TeV before suffering a drop-off.

In contrast, in frame {\it b}) we plot the distribution of third generation scalar masses $m_0(3)$. 
In this case, for $n=1,2$ the distribution peaks around 5-6 TeV while dropping to near zero around
10 TeV for $n=1$ and 12 TeV for $n=2$. Large values of $m_0(3)$ generate large stop masses which result in $\Sigma_u^u(\tst_{1,2})$ 
exceeding $\sim 30$ {\it i.e.} generating a weak scale typically in excess of $m(weak)\sim 400$ GeV.
For $n=0$, the distribution peaks around 3 TeV.

In frame {\it c}), we plot the distribution in $m_{1/2}$. In this case, the $n=1$ distribution peaks around
1.5 TeV whilst $n=2$ peaks slightly higher. If the (unified) gaugino masses become too big, then
the large gluino mass also lifts the top squarks to higher masses thus causing the $\Sigma_u^u(\tst_{1,2})$
to again become too large. The distributions fall to near zero by $m_{1/2}\sim 3$ TeV leading to
upper limits on gaugino masses. The $n=0$ distribution actually peaks at its lowest allowed values
followed by a steady decline. 

In frame {\it d}), we show the distribution versus $A_0$. Here we only show the more lucrative
negative $A_0$ case which leads to higher Higgs masses $m_h$\cite{h125}. 
The $n=0$ distribution peaks at $A_0\sim 0$ with a steady fall-off at large negative $A_0$ values. 
In this case, the typically small mixing in the stop sector leads to values of $m_h$ below its measured
result.
In contrast, 
for $n=1,2$ the distributions increase (according to the statistical pull) to peak values
around $A_0\sim -(5-10)$ TeV. 
Such large $A_0$ values lead to large mixing in the top-squark
sector which can enhance $m_h$ whilst decreasing the $\Sigma_u^u(\tst_{1,2})$ values\cite{ltr}.
The $n=1$ curve actually features a double bump structure: we have traced the lower peak to
the presence of large $m_A\sim m_{H_d}\sim 5-10$ TeV values which increase the $S$ term in the
third generation matter scalar RGEs. This term (along with large two-loop effects from
first/second generation matter scalars) acts to suppress $m_{U_3}^2$ leading to lighter 
$\tst_1$ states even without large mixing. 
For even larger negative $A_0$ values, the distributions rapidly fall to zero since they start
generating CCB minima in the MSSM scalar potential.

\begin{figure}[t]
  \centering
  {\includegraphics[width=.48\textwidth]{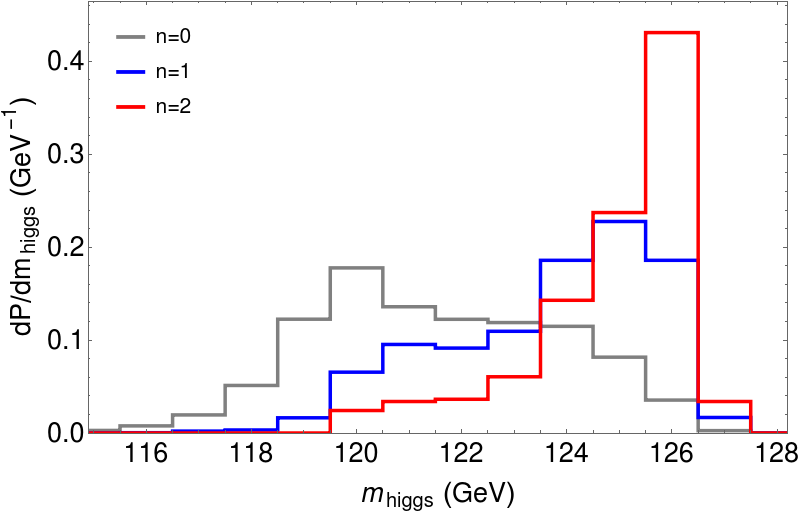}}\quad
  {\includegraphics[width=.48\textwidth]{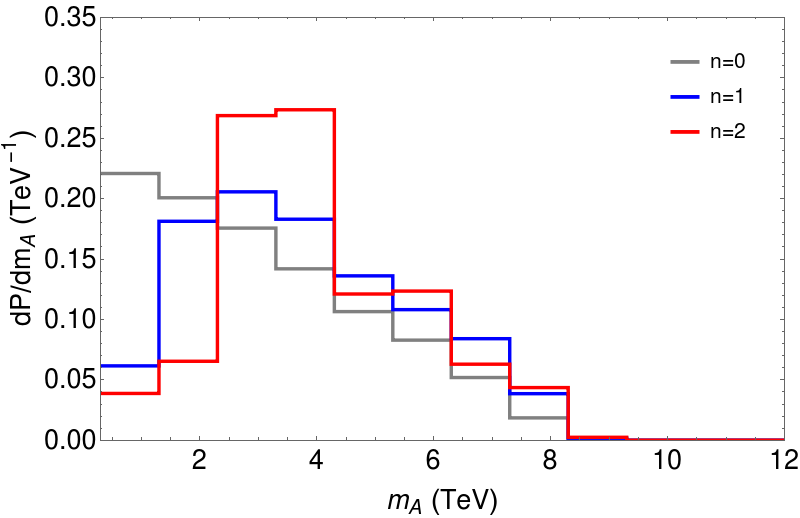}}\\ 
  {\includegraphics[width=.48\textwidth]{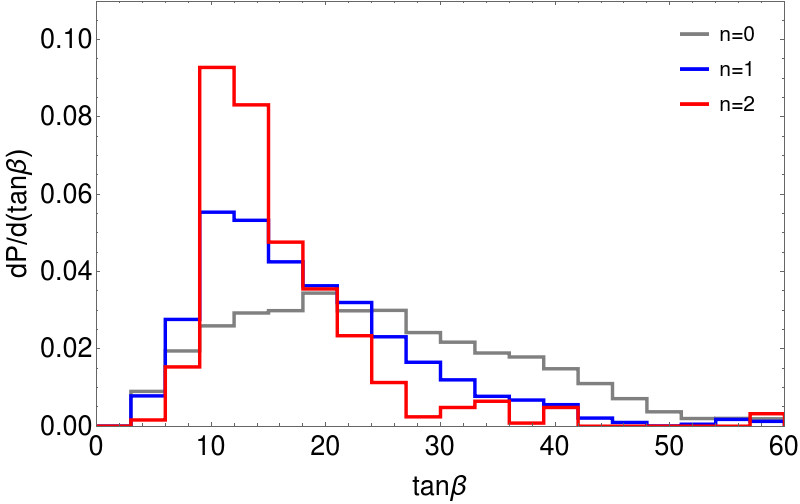}} \quad
  {\includegraphics[width=.48\textwidth]{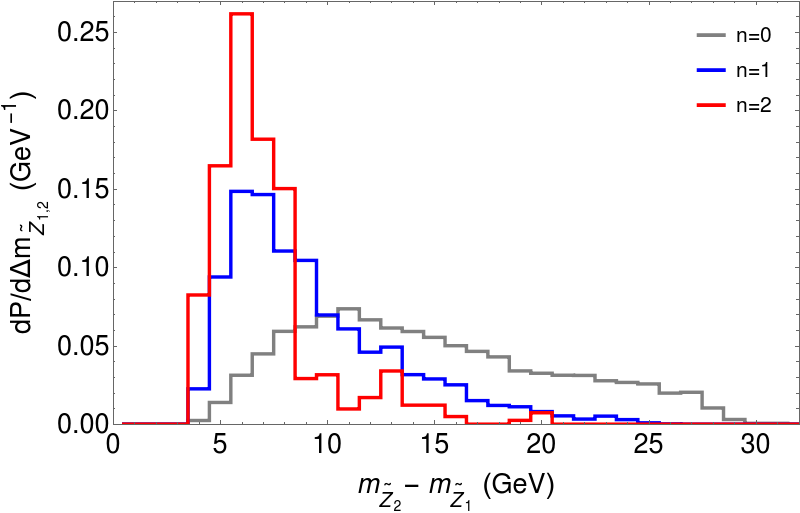}}\\
  \caption{Case\ {\bf A}:\ \ \ $f_{EWFT}\to \Theta (30 -\Delta_{\rm EW})$ - Upper panels: 
Probability distributions in $m_h$ (left) and $m_{A}$  (right). 
Lower panels: Probability distributions in $\tan\beta$  (left) and $m_{\tz_2}-m_{\tz_1}$ (right). 
All distributions are shown following Eq.~\ref{fullmeasure} and in addition 
rejecting CCB and noEWSB minima. 
Results for different values of $n = 2n_F+n_D -1$ are displayed for each plot. 
  }  
  \label{fig:3}
\end{figure}

In Fig. \ref{fig:3}, we show string landscape probability predictions for quantities associated 
with the Higgs and electroweak-ino sector. 
Special attention should be paid to the Higgs mass distributions. 
In frame {\it a}), we show $dP/dm_h$ vs. $m_h$ for 
$n=0,\ 1$ and $2$. 
For $n=0$, we find a broad peak ranging from $m_h\sim 119-125$ GeV. 
This may be expected for the $n=0$ case since we have a uniform scan in soft terms 
and low $\Delta_{\rm EW}$ can be found for
$A_0\sim 0$ which leads to little mixing in the stop sector and hence too light values of $m_h$. 
Taking $n=1$, instead we now see that the distribution in $m_h$ peaks at $\sim 125$ GeV with the bulk
of probability between $123$ GeV $<m_h<$127 GeV-- in solid agreement with the measured value of 
$m_h=125.09\pm 0.24$ GeV\cite{pdg}.\footnote{Here, we rely on the Isajet 7.87 theory evaluation of $m_h$
which includes renormalization group improved 1-loop corrections to $m_h$ along with leading 
two-loop effects. Calculated values of $m_h$ are typically within 1-2 GeV of similar calculations from
latest FeynHiggs\cite{feynhiggs} and SUSYHD\cite{susyhd} codes.} This may not be surprising since the landscape is pulling the various 
soft terms towards large values including large mixing in the Higgs sector which lifts up $m_h$ into the
125 GeV range. By requiring the $\Sigma_u^u(\tst_{1,2})\alt 30$ 
(which would otherwise yield a weak scale in excess of 350 GeV) then too large of Higgs masses are vetoed. For the $n=2$ case with a stronger draw towards
large soft terms, the $m_h$ distribution hardens with a peak at $m_h\sim 126$ GeV.

In Fig. \ref{fig:3}{\it b}), we show the distribution in pseudoscalar mass $m_A$. 
Here, for $m_A\gg m_h$, then $m_A\sim m_{H_d}$ (at the weak scale) 
and we have a statistical draw to large $m_A$ values which is tempered by the presence of
$m_{H_d}/\tan\beta$ in Eq. \ref{eq:mzs}. While the $n=0$ uniform draw peaks at the lowest $m_A$ values, 
the $n=1$ and 2 cases yield a broad distribution peaking around $m_A\sim3$ TeV which drops thereafter.
In frame {\it c}), we show the distribution in $\tan\beta$. Here, the $n=0$ case has a broad distribution with 
a peak around $\tan\beta\sim 20$ while the $n=1$ and 2 cases have sharper distributions peaking around
$\tan\beta\sim 10-15$. The suppression of $\tan\beta$ for large values can be understood due to the
draw towards large soft terms in the sbottom sector. As $\tan\beta$ increases, the $b$ (and $\tau$) Yukawa couplings
increase so that the $\Sigma_u^u(\tb_{1,2})$ terms become large. Then the anthropic cutoff on $\Delta_{\rm EW}<30$
disfavors the large $\tan\beta$ regime. In frame {\it d}), we show the $m_{\tz_2}-m_{\tz_1}$ mass splitting.
For our case with $\mu=150$ GeV, the light higgsinos $\tw_1^\pm$, $\tz_{1,2}$ all have masses around 150 GeV.
The phenomenologically important mass gap $m_{\tz_2}-m_{\tz_1}$ becomes smaller the more gauginos are 
decoupled from the higgsinos. The landscape draw towards large gaugino masses thus suppressed 
$m_{\tz_2}-m_{\tz_1}$ for the $n=1$ and $2$ cases so that the mass gap peaks at around $5-8$ GeV.
For the uniform scan with $n=0$, then the gap is larger-- typically $10-20$ GeV.

\begin{figure}[t]
  \centering
  {\includegraphics[width=.48\textwidth]{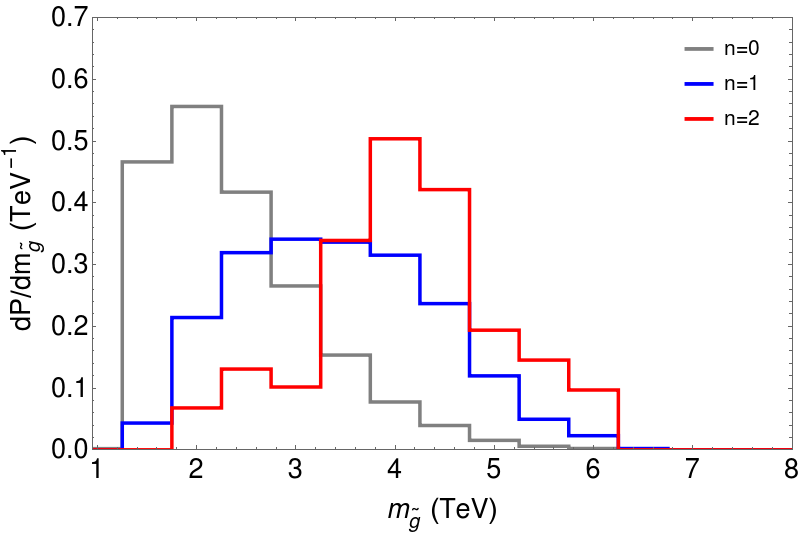}}\quad
  {\includegraphics[width=.48\textwidth]{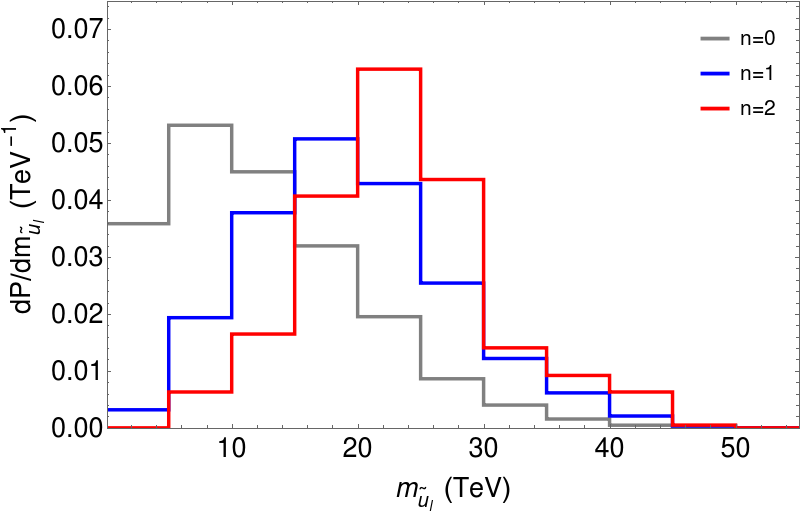}}\\ 
  {\includegraphics[width=.48\textwidth]{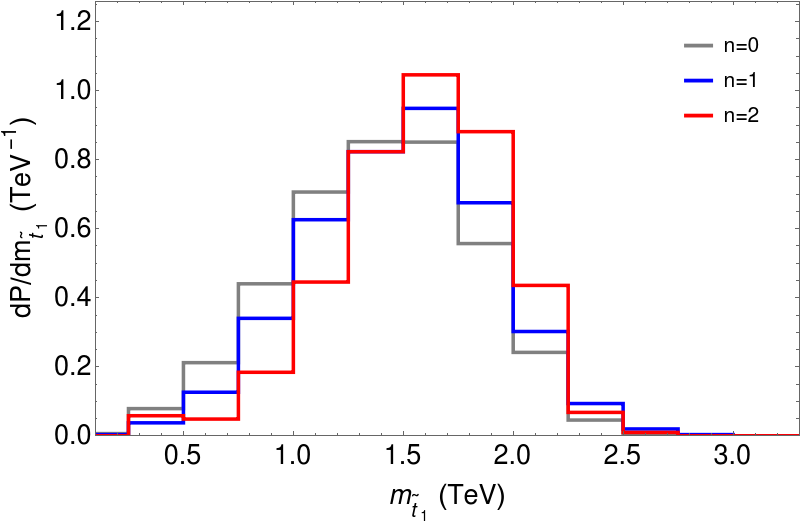}} \quad
  {\includegraphics[width=.48\textwidth]{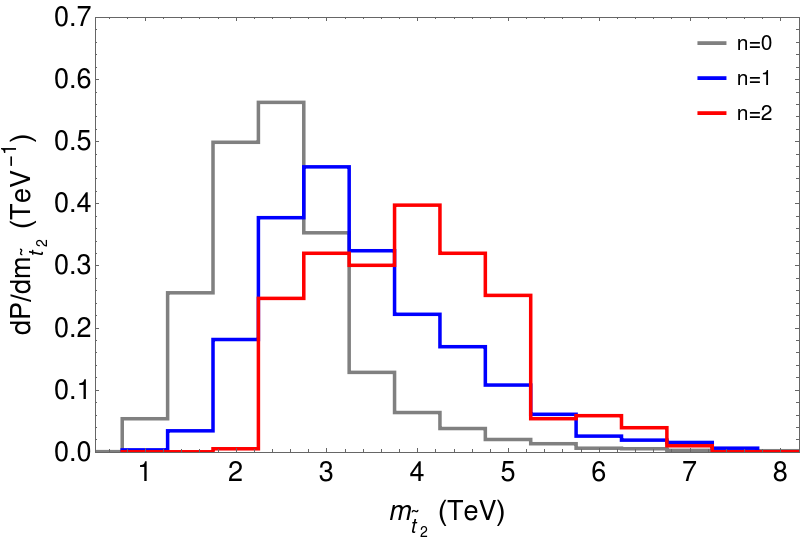}}\\
  \caption{Case\ {\bf A}:\ \ \ $f_{EWFT}\to \Theta (30 -\Delta_{\rm EW})$ - 
Upper panels: Probability distributions in $m_{\tg}$ (left) and $m_{\tu_L}$  (right). 
Lower panels: Probability distributions in $m_{\tst_1}$  (left) and $m_{\tst_2}$ (right). 
All distributions are shown following Eq.~\ref{fullmeasure} and in addition rejecting 
CCB and noEWSB minima. 
Results for different values of $n = 2n_F +n_D-1$ are displayed for each plot. 
  }  
  \label{fig:4}
\end{figure}

In Fig. \ref{fig:4} we show string landscape probability distributions for some 
strongly interacting sparticles. In frame {\it a}), we show the distribution in gluino mass
$m_{\tg}$. From the figure, we see that the $n=1$ distribution rises to a peak probability
around $m_{\tg}=3.5$ TeV. This may be compared to current LHC13 limits which require
$m_{\tg}\agt 2$ TeV\cite{lhc_mgl}. 
Thus, it appears LHC13 has {\it only begun} to explore the relevant
string theory predicted mass values. 
The distribution fall steadily such that essentially 
no probability exists for $m_{\tg}\agt 6$ TeV. 
This is because such heavy gluino masses lift the top-squark sector soft terms under RG running
so that $\Sigma_u^u(\tst_{1,2})$ then exceeds 30.
For $n=2$, the distribution is somewhat harder, peaking at around $m_{\tg}\sim 4.5$ TeV.
The uniform $n=0$ distribution peaks around 2 TeV.

In frame {\it b}), we show the distribution versus one of the first generation
squark masses $m_{\tu_L}$. Here, it is found for $n=1,\ 2$ that the distribution peaks
around $m_{\tq}\sim 20-25$ TeV-- well beyond LHC sensitivity, but in the range to provide at 
least a partial decoupling solution to the SUSY flavor and CP problems.
It would also seem to reflect a rather heavy gravitino mass $m_{3/2}\sim 10-30$ TeV in accord 
with a decoupling solution to the cosmological gravitino problem\cite{linde}. The $n=0$ 
distribution peaks around $m_{\tq}\sim 8$ TeV and drops steadily to the vicinity of 40 TeV.
For much heavier squark masses, then two-loop RGE terms tend to drive the stop sector 
tachyonic resulting in CCB minima.

In frame {\it c}), we show the probability distribution versus $m_{\tst_1}$. In this case, 
all three $n$ values lead to a peak around $m_{\tst_1}\sim 1.5$ TeV. 
While this may seem surprising at first, in the case of $n=1,\ 2$ we gain large $A_t$ trilinear terms 
which lead to large mixing and a diminution of the eigenvalue $m_{\tst_1}$\cite{ltr} even though the soft terms
entering the stop mass matrix may be increasing. There is not so much probability below 
$m_{\tst_1}=1$ TeV which corresponds to recent LHC13 mass limits\cite{lhc_mt1}. 
Thus, again, LHC13 has only begun to explore the predicted string theory parameter space.
The distributions taper off such that hardly any probability is left beyond 
$m_{\tst_1}\sim 2.5$ TeV. This upper limit is apparently within reach of high-energy LHC
operating with $\sqrt{s}\sim 27$ TeV where the reach in $m_{\tst_1}$ 
extends to about $2.5-3$ TeV\cite{lhc33_2}.
In frame {\it d}), we show the distribution in $m_{\tst_2}$. In this case, the suppression
of $m_{\tst_2}$ from large mixing $A_t$ is far less and so the $n=1,\ 2$ distributions peak at higher values $m_{\tst_2}\sim 3-5$ TeV as compared to the uniform $n=0$ scan where 
$m_{\tst_2}$ peaks around 2 TeV. The distributions fall steadily so that hardly any probability
exists beyond $m_{\tst_2}\agt 6$ TeV because the $\Sigma_u^u(\tst_2 )$ 
values become too large.

Let us summarize  our main conclusions from this Section. 
We find that the anthropic requirement of a weak scale not too removed (by a factor 4) from
its measured value (which is imposed by requiring $\Delta_{\rm EW} \leq 30$) 
centers the low-energy supersymmetric spectrum around central values that are 
relatively agnostic about the precise distribution of supersymmetry breaking scales in the UV
so long as $n\ge 1$. 
There is some shift in the predicted supersymmetric spectrum as 
$n = 2 n_F +n_D- 1$ is varied, but the shift is relatively minor.

The $n=0$ case we regard as rather implausible compared to
$n=1,\ 2$ in that it typically generates $m_h<123$ GeV (allowing for a couple GeV 
theory error in our $m_h$ calculation).
It is intriguing that the best prediction for $m_h\sim 125$ GeV is obtained with $n=1$ 
which corresponds to SUSY breaking dominated by a single auxiliary field $F$, a situation
that is rather common in the literature. 

\subsubsection{Cases with $n\ge 4$}

We have also tried a case with $n=4$. 
In that case, the soft term generation became extremely inefficient 
since almost always one is placed into either CCB or no EWSB 
vacua or else $\Delta_{\rm EW}\gg 30$. 
This may be understood from examining Fig. 1 of Ref. \cite{lscape}.
If the $A_0$ parameter is generated at too large values compared to $m_0(3)$,
then the $m_{{\tilde t}_R}^2$ soft term gets driven to negative values at the 
weak scale resulting in CCB minima for the scalar potential. If $m_{H_u}^2$ 
is generated at too large values, then it isn't even driven negative so that
electroweak symmetry isn't properly broken. 

\begin{figure}[t]
  \centering
  {\includegraphics[width=.7\textwidth]{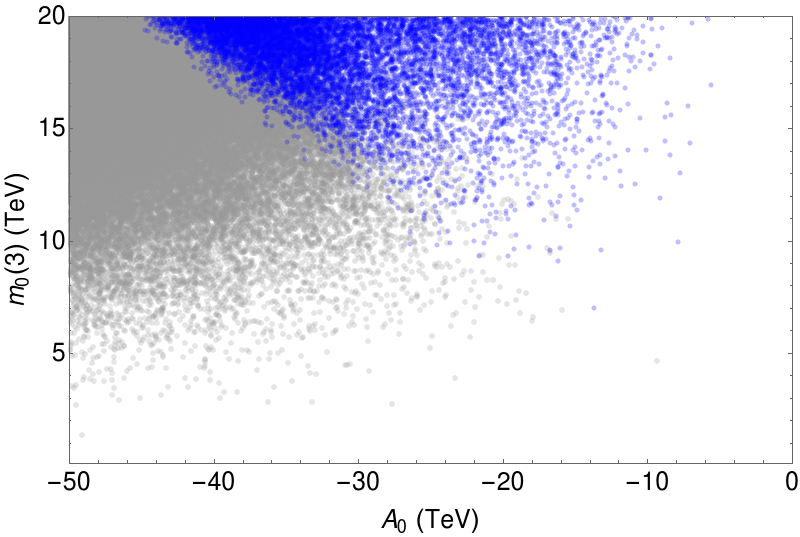}}\quad
  \caption{Locus of 100K scan points from a scan with $n=4$ and scan range as below Eq. \ref{eq:m32}. 
The gray points have either CCB scalar potential minima or no
EWSB. 
The blue points admit EWSB but all have $\Delta_{\rm EW}>240$ 
corresponding to a weak scale greater than $\sim 1$ TeV.
}  
\label{fig:n=4}
\end{figure}

The situation is illustrated in Fig.~\ref{fig:n=4} where we plot the locus of 
$n=4$ scan points using the scan limits below Eq. \ref{eq:m32}. 
We show for clarity just 100K points although we have generated 1M.
The large value of $n$ selects almost always huge values of soft terms which
then either lead to invalid scalar potential minima or else, 
if EW symmetry is properly broken, a huge value for the weak scale 
due to huge values of $\Sigma_u^u(i)$ or $-m_{H_u}^2(weak)$. 
The large $n$ scenario only gets worse if we increase the 
(artificial) scan  upper limits from below Eq.~\ref{eq:m32}.
This may be an important result for string model builders
in that $n\ge 4$ is difficult to accommodate phenomenologically: 
realistic vacua with the weak scale $m_{W,Z,h}\sim 100$ GeV 
seem to prefer $n\sim 1-2$.

\subsubsection{Varying the $\Delta_{\rm EW}$ cutoff}

\begin{figure}[t]
  \centering
  {\includegraphics[width=.7\textwidth]{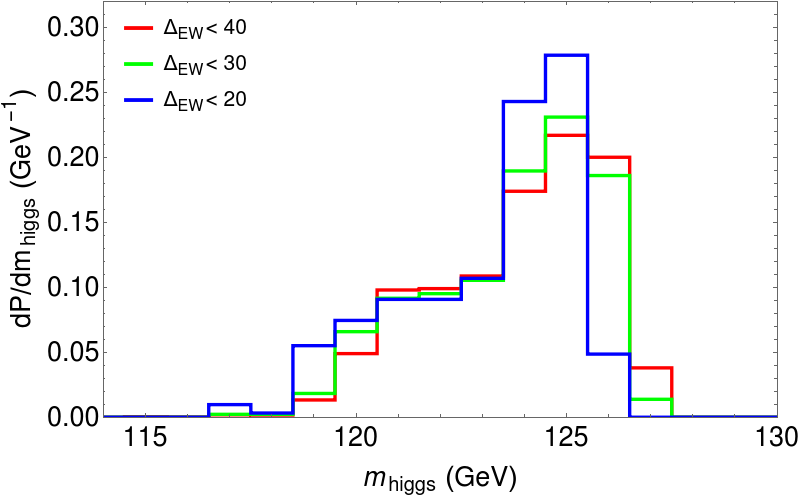}}\quad
  \caption{Probability distribution for Higgs mass $m_h$ for the case of $n=1$ but with
varying cutoff $\Delta_{\rm EW}<20$, 30 and 40.
  }  
  \label{fig:mh_dew}
\end{figure}

What happens if we vary the cutoff for $\Delta_{\rm EW}$? 
In Fig. \ref{fig:mh_dew} we show the probability distribution 
for the Higgs mass $m_h$ for $n=1$ but for three choices
of cutoff $\Delta_{\rm EW}<20$, 30 and 40. From the distributions, we see that the
$m_h$ distributions slightly hardens with an increasing cutoff but overall 
$m_h\sim 124-126$ GeV is still predicted. In the next Subsection we explore 
what happens using instead the case {\bf B} prescription for $f_{EWFT}$.

\subsubsection{Conclusions for case {\bf A}:}

It would thus appear that when statistical questions of distributions 
in the landscape are tempered with anthropic requirements, 
more or less solid predictions about the IR spectrum are obtained. 
We also note our other main conclusion-- 
the mass of the Higgs comes out close to its observed value--  
is robust against variations in $n = 1$ or $2$
and also against variations in the cutoff value of $\Delta_{\rm EW}$.

\subsection{Case {\bf B}:}

In this Subsection, we examine the results of our numerical scans using 
$f_{SUSY}\sim m_{soft}^n$ but now with $f_{EWFT}=\Delta_{\rm EW}^{-1}$. In this case we 
can veto parameter space points statistically according to a $\Delta_{\rm EW}(min)/\Delta_{\rm EW}$ 
algorithm or else bin surviving events with a variable weight given by the same factor.
In either case, the surviving weights will be penalized by a factor $\Delta_{\rm EW}^{-1}$. 
Although such a factor penalizes events with a large computed weak scale, it does 
nonetheless allow many to survive. The question is: is the penalty sufficient to offset the
$f_{SUSY}\sim m_{soft}^n$ draw towards large soft terms for $n\ge 1$. 

\begin{figure}[t]
  \centering
  {\includegraphics[width=.3\textwidth]{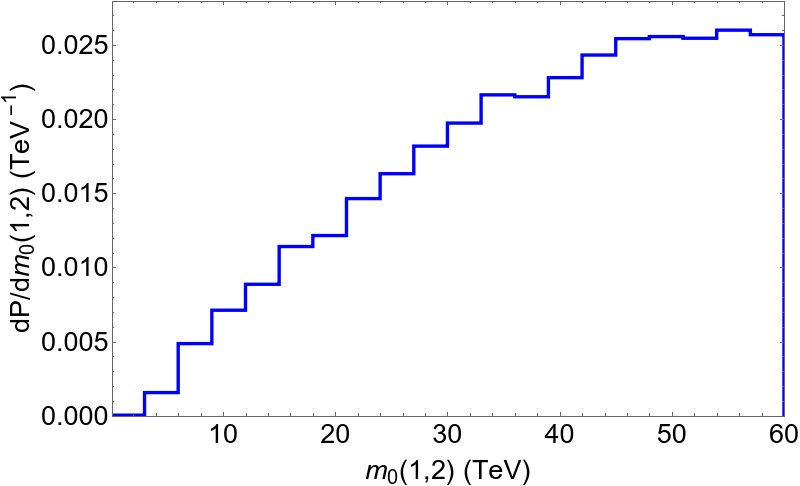}}\quad
  {\includegraphics[width=.3\textwidth]{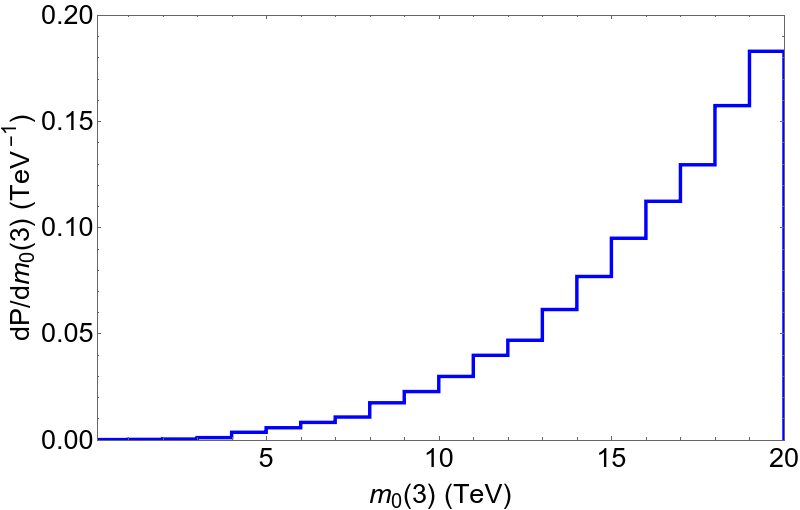}}\quad
  {\includegraphics[width=.3\textwidth]{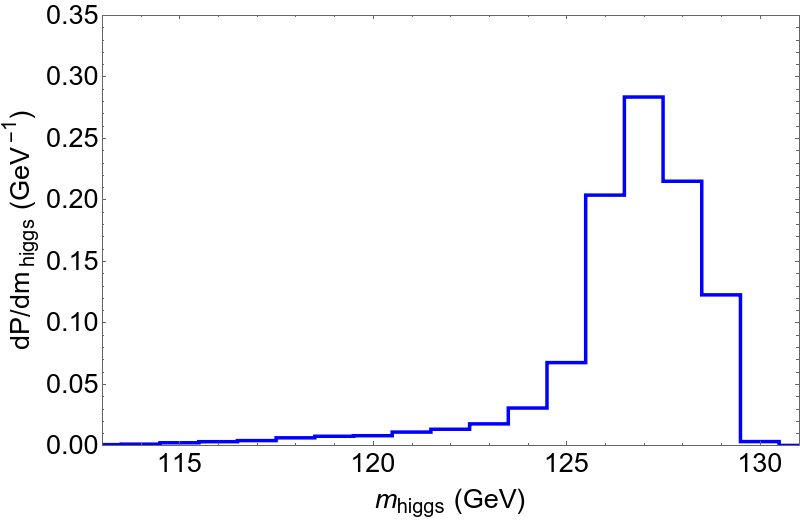}} \quad 
  \caption{Case\ {\bf B}:\ \ \ $f_{EWFT}\to \Delta_{\rm EW}^{-1}$: Distributions in {\it a}) $m_0(1,2)$, {\it b}) $m_0(3)$ 
and {\it c}) $m_h$ for $n=1$. 
  }  
  \label{fig:5}
\end{figure}

In Fig. \ref{fig:5}, we show our first results from Case {\bf B}. 
We scan over the same soft parameter ranges as in case {\bf A}.
In frame {\it a}) ({\it b})), we see the probability distribution of vacua versus first/second generation
matter scalar soft masses $m_0(1,2)$ (third generation soft masses $m_0(3)$). 
For these cases, the $\Delta_{\rm EW}^{-1}$ penalty is insufficient to create an upper bound on matter scalar
masses and hence the upper bounds come merely from our scan limits above. 
For this case, in frame {\it c}) we show the vacua probability versus $m_h$. 
Here, the value of $m_h\sim 126-129$ GeV which is a reflection of the rather high values of the soft 
terms which are allowed. For case {\bf B}, the penalty $\Delta_{\rm EW}^{-1}$ allows for events with 
far higher values of $m_{weak}$ in the TeV range, well beyond the $\sim 100$ GeV value. 

\begin{figure}[t]
  \centering
  {\includegraphics[width=.3\textwidth]{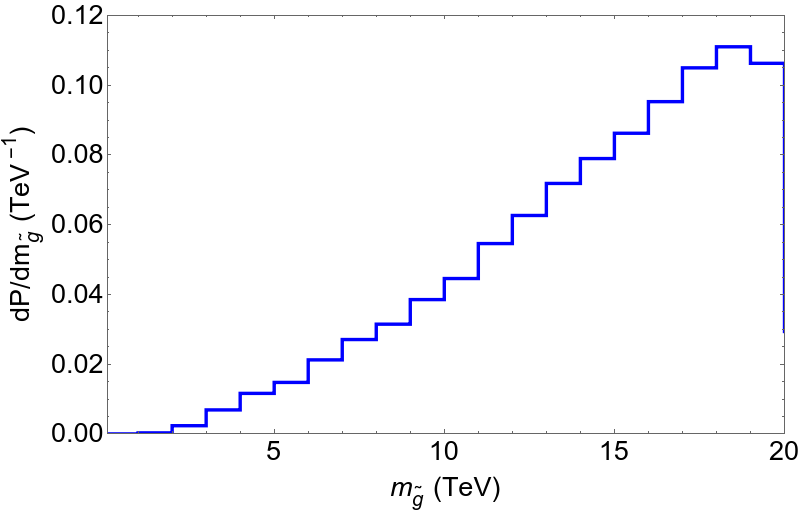}}\quad
  {\includegraphics[width=.3\textwidth]{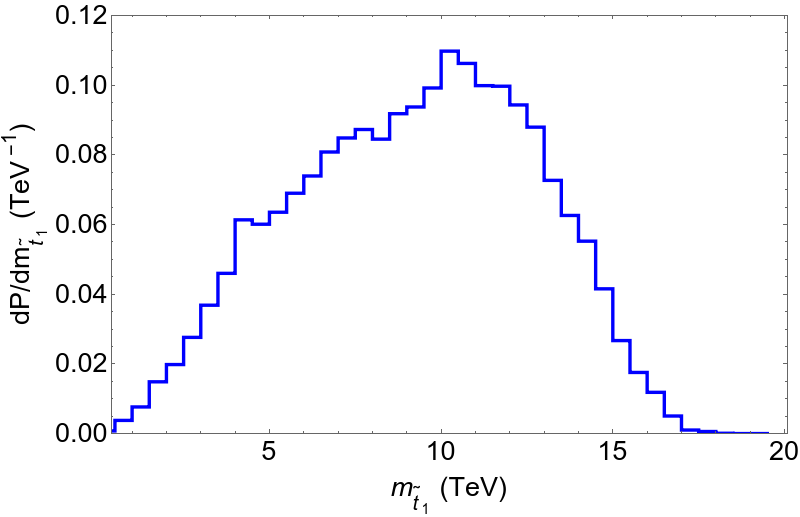}}\quad
  {\includegraphics[width=.3\textwidth]{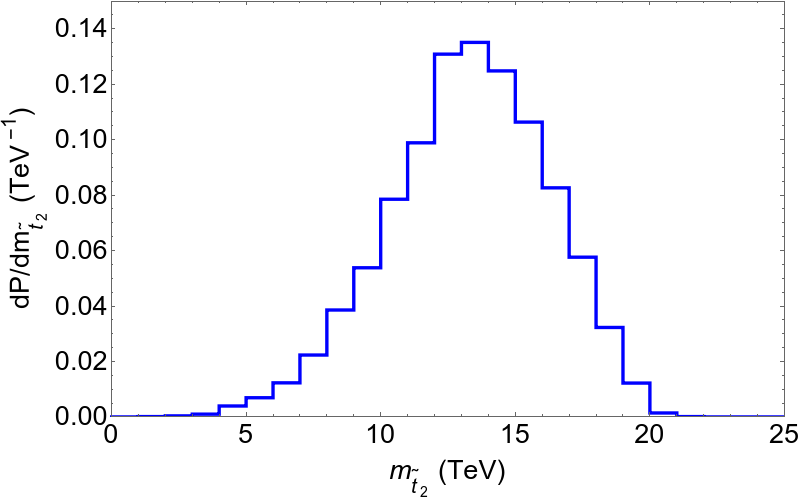}} \quad 
  \caption{Case\ {\bf B}:\ \ \ $f_{EWFT}\to \Delta_{\rm EW}^{-1}$: Distributions in {\it a}) $m_{\tg}$, {\it b}) $m_{\tst_1}$ 
and {\it c}) $m_{\tst_2}$ for $n=1$.
  }  
  \label{fig:6}
\end{figure}

In Fig. \ref{fig:6}, we show distributions in {\it a}) $m_{\tg}$, {\it b}) $m_{\tst_1}$ and
{\it c}) $m_{\tst_2}$ from the case {\bf B} scan. We see that much higher mass scales are 
favored due to allowing much higher values of $m_{weak}$. 
In particular, here values of $m_{\tg}\sim 20$ TeV,   
$m_{\tst_1}\sim 10$ TeV and $m_{\tst_2}\sim 14$ Tev are favored. 
In this case, the spectra has clearly entered the unnatural region and 
so we do not pursue the case {\bf B} avenue any further.

\section{Implications for collider and dark matter searches}
\label{sec:col_dm}

\subsection{Colliders:}

Here we will focus on our case {\bf A} results with $n=1$ or 2 since these results predict a Higgs boson
mass very close to or at its measured value. In this case, we may wish to take the remaining 
sparticle mass predictions seriously as well. As far as LHC searches go, we have found from Fig. \ref{fig:4}
that there is only a tiny probability that $m_{\tg}$ lies below the $m_{\tg}>2$ TeV mass bound. 
This means LHC has only begun to explore the string theory parameter space. 
Recently, the reach of HL-LHC (high luminosity LHC with $\sqrt{s}=14$ TeV and $\sim 3$ ab$^{-1}$ of 
integrated luminosity) has been estimated for gluinos\cite{mgluino} and for top squarks\cite{atlas,stop}:
it extends at $5\sigma$ level to $m_{\tg}\sim 2.8$ TeV and $m_{\tst_1}\sim 1.4$ TeV. 
Thus, from Fig. \ref{fig:4} we see that there is a large probability that SUSY would 
escape HL-LHC searches in the gluino pair or top-squark pair production channels. 
However, the HE-LHC (high energy LHC with $\sqrt{s}=27$ TeV and $10-15$ ab$^{-1}$)
has a reach extending to $m_{\tg}\sim 5.5$ TeV\cite{lhc33_1} and $m_{\tst_1}\sim 3$ TeV\cite{lhc33_2}. 
This should be enough to cover the probability distributions in Fig. \ref{fig:4}. 

Of relevance for HL-LHC searches is the same sign diboson
SUSY discovery channel arising from charged/neutral wino pair production in models with light higgsinos\cite{ssdb}:
$pp\to\tw_2\tz_4\to (W^\pm\tz_{1,2})+(\tw_1^\mp W^\pm)$ where the heavier higgsinos are quasi-visible due to their
low visible energy release and the lightest higgsino $\tz_1$, which comprises a portion of dark matter, is
completely invisible. The HL-LHC reach in this channel is to $m_{\tw_2}\sim 1$ TeV corresponding roughly to
$m_{1/2}\sim 1.2$ TeV. Again, this covers only a portion of string parameter space from Fig. \ref{fig:2}{\it c}).

A final LHC SUSY discovery channel\cite{lljMET,cms} arises from direct higgsino 
pair production $pp\to\tz_1\tz_2+jet$ with $\tz_2\to\tz_1\ell^+\ell^-$.\footnote{
A related channel is monojet production from $pp\to\tz_1\tz_1j$ production yielding a
$jet+\eslt$ signature from initial state radiation recoiling against the two WIMPs. 
This channel has been investigated in Ref. \cite{azar} where the signal is found to occur at the 
1\% level compared to SM background from $pp\to Zj$ production with $Z\to\nu\bar{\nu}$ and
where signal and BG have very similar $\eslt$ and $p_T(jet)$ distributions. 
Thus, the monojet channel does not seem to be a viable discovery channel for SUSY.} 
This challenging channel is potentially most powerful for SUSY models 
with light higgsinos although in our case from Fig. \ref{fig:3}{\it d}) the expected 
$m_{\tz_2}-m_{\tz_1}$ mass gap is expected to occur in the $4-8$ GeV range so the dilepton pair will
occur with very low $p_T$ values\footnote{We refer to \cite{Delannoy:2013ata} for some additional 
LHC studies conducted in this direction.}. 

Of course, a higher energy $e^+e^-$ collider operating with $\sqrt{s}>2m(higgsino)$ would be able to
cover all parameter space and indeed would then function as a {\it higgsino factory}\cite{ilc}. 
In our case, with $\Delta_{\rm EW}<30$, this corresponds to higgsino masses below about
350 GeV so a machine such as ILC with $\sqrt{s}\sim 500-700$ GeV may be needed.

\subsection{Dark matter searches:}

For all of our discussion, we have assumed a weak scale $m_{weak}\alt 350$ GeV which corresponds 
to $\mu\alt 350$ GeV so that the lightest higgsino is the lightest SUSY particle and constitutes a 
portion of dark matter. 
If some mechanism such as radiative PQ breaking generates the $\mu$ parameter, 
as discussed in Sec. \ref{sec:intro}, then the remainder of dark matter would be a SUSY DFSZ axion\cite{bbc}. 
Calculations of the mixed axion/higgsino dark matter relic density typically predict the bulk of DM to 
lie in axions (typically 80-90\%)  while 10-20\% lies in higgsino-like WIMPs\cite{Bae:2014rfa}. 
Nonetheless, prospects for WIMP detection are 
good at ton-scale noble liquid detectors even though the WIMP target abundance is typically well below
that which is usually assumed. Detailed calculations show multi-ton WIMP detectors should cover 
all of parameter space\cite{hwimps}. 

\begin{figure}[t]
  \centering
  {\includegraphics[width=.7\textwidth]{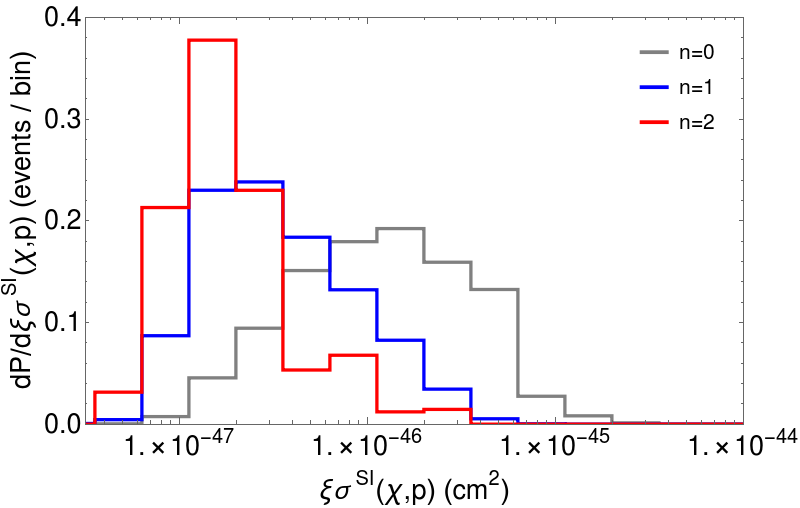}}\quad
  \caption{Plot of $dP/d\xi\sigma^{SI}(\tz_1 p)$ versus $\xi\sigma^{SI}(\tz_1 p)$ for
case {\bf A} scans with $n=0$, 1 and 2.
}  
\label{fig:SI}
\end{figure}

In Fig. \ref{fig:SI}, we show the distribution $dP/d\xi\sigma^{SI}(\tz_1,p )$ versus $\xi\sigma^{SI}(\tz_1,p )$ 
for various $n$ values. Current limits from LUX\cite{lux} and PandaX\cite{pandax} 
require $\xi\sigma^{SI}(\tz_1 p)<2\times 10^{-46}$ cm$^{2}$ for $m_{\tz_1}\sim 150$ GeV. The quantity 
$\xi\equiv \Omega_{\tz_1}^{TP}h^2/0.12$ measures the minimal fraction of dark matter as composed of thermally-produced WIMPs rather than axions and is typically $0.05-0.1$ for mixed light higgsino/axion dark matter. While about half the 
parameter space seems explored by ton-scale WIMP detectors for the uniform scan with $n=0$, the distribution
skews to lower values as $n$ increases to $1$ or 2. This is because as $n$ increases, the 
gaugino masses are drawn to larger values while $\mu$ remains fixed and the $\tz_1$ becomes more purely higgsino-like.
The $\tz_1-\tz_1-h$ coupling is a product of gaugino times higgsino components\cite{hwimps} so 
typically decreases as the gaugino-higgsino mass gap increases. Only a small portion of parameter
space is ruled out for $n=1$ or $2$ although future probes down to $\sim 10^{-47}$ cm$^2$ will cover just about all parameter space.

\begin{figure}[t]
  \centering
  {\includegraphics[width=.7\textwidth]{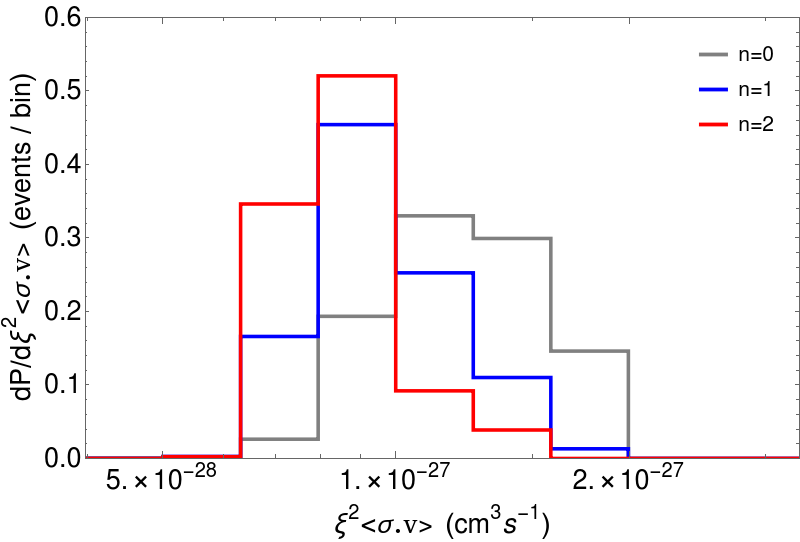}}\quad
  \caption{Plot of $dP/d\xi^2\langle\sigma v\rangle$ versus 
$\xi^2\langle\sigma v\rangle$ for case {\bf A} scans with $n=0$, 1 and 2.
}  
\label{fig:IDD}
\end{figure}

In Fig. \ref{fig:IDD} we show the distribution in $dP/d\xi^2\langle\sigma v\rangle$ vs. $\xi^2\langle\sigma v\rangle$ for the case {\bf A} scans with
$n=0$, 1 and 2. Here the $\xi$ factor is squared due to the necessity of having
indirect detection of WIMP-WIMP annihilation in the cosmos. The best limits for
$m_{\tz_1}\sim 150$ GeV come from Fermi-LAT/MAGIC combined limits\cite{fermi} on
 observation of gamma rays from dwarf spheroidal galaxies; these require
$\xi^2\langle\sigma v\rangle <3\times 10^{-26}$ cm$^3$/s. 
As can be seen, all predictions are well below the limits 
due partly to the depleted WIMP abundance squared. As $n$ increases, 
the detection rates drop due to increasing sparticle masses which suppress
the WIMP annihilation cross section.

In the case of axions, the SUSY DFSZ axion coupling to photons has been 
found to be severely diminished (by about an order of magnitude) compared to expectations 
from non-SUSY models due to the presence of light higgsinos in the axion-$\gamma$-$\gamma$ triangle diagram\cite{susyaxion}.
Thus, axion detectors which probe much more deeply into small $a\gamma\gamma$ 
coupling strengths will be needed.

\section{The Cosmological Moduli Problem}
\label{sec:cosmo}

We have seen in the previous sections that introducing anthropic constraints on the landscape had two kinds of effects on the low energy supersymmetric spectrum: $(i)$ for the vacuum energy, the constraint did not affect the selection of our supersymmetry breaking vacuum; $(ii)$ for the electroweak scale, the constraint had the effect of selecting natural values of the superpartner masses.

A generic issue that affects the kind of arguments we have presented here is the cosmological moduli problem (originally the Polonyi problem, dating from the earliest theories of supergravity\cite{Coughlan:1983ci}). The energy density of the Universe can be dominated by moduli fields, which, being gravitationally coupled to matter, can decay at late times. 
If the lifetime of moduli exceeds the era of Big Bang Nucleosynthesis, 
then late decay of moduli can disassociate the newly created nuclei and 
ruin the successful prediction of abundances of the light elements. 

Over the last decade and a half, significant progress has been made on the issue of moduli stabilization in string theory \cite{Douglas:2006es}. Most moduli acquire masses near the string scale from a combination of effects - fluxes, branes, and strong coupling in the hidden sector. However, one also generally expects moduli which are parametrically lighter than the string scale, 
and satisfy\cite{Acharya:2010af} 
\be
m_{modulus} \, \sim \, m_{3/2} \,\,\,\,. 
\ee
Such light moduli decay around $t \sim M^2_{P}/m^3_{modulus} \sim 10^3$ s for $m_{modulus} \sim 1$ TeV. This clearly interferes with the successful predictions of Big Bang Nucleosynthesis \cite{Kane:2015jia}, \cite{Dutta:2009uf}. An equivalent way to express this is in terms of the reheat temperature
\be
T_r \, = \, c^{1/2} \left( \frac{10.75}{g_*} \right)^{1/4} \left( \frac{m_{modulus}}{\rm 50~ TeV} \right)^{3/2} T_{BBN}
\ee
where the decay width is given by $\Gamma  = \frac{c}{2 \pi} \frac{m^3_{modulus}}{M^2_P}$.

To avoid conflicts with Big Bang Nucleosynthesis, one thus typically requires
\be
m_{modulus} \, \geq \, 50 \,\,{\rm TeV} \,\,.
\ee
From the point of view of  distributions of permissible vacua, this would introduce a biasing factor
\be \label{fullmeasurewithmoduli}
dN_{vac} \, \sim \,  \Theta (m_{modulus} - 50 \,\,{\rm TeV}) \times f_{EWFT} \times (m^2_{hidden})^n d(m^2_{hidden})  \,\,
\ee
Now, using the fact that $m_{moduli }\sim m_{3/2}$, and the relations between the gravitino mass and soft terms from Eq.~\ref{eq:m32}, we can recast the condition of avoiding the cosmological moduli problem as
\be \label{fullmeasureonsofts}
d N_{vac} \, \sim \,  \Theta (m_{0} - c_1*50 \,\,{\rm TeV}) \times \Theta (m_{1/2} - c_2*50 \,\,{\rm TeV}) \times    f_{EWFT} \times (m^2_{hidden})^n d(m^2_{hidden})  \,\,.
\ee

We then see that there are two opposing tendencies here. The pull to natural solutions, embodied by the $ f_{EWFT}$ term, is opposed to the pull for vacua where 
the moduli problem is avoided, which are the origin of the first two step functions in Eq.~\ref{fullmeasureonsofts}. Indeed, in our scan, we specifically imposed 
upper limits $m_0(1,2) < 60$ TeV, $m_0(3) < 20$ TeV, and $m_{1/2} < 10$ TeV. 
This was in anticipation of the fact that solutions beyond the imposed upper limits would be 
cut off by the requirement on $\Delta_{\rm EW}$, leading to inefficient scanning. 
However, these larger values of the soft terms turn out to be precisely the ones 
needed to solve the moduli problem.

In our opinion, this points to the fact that regions of the landscape where the coefficients $c_1$ and $c_2$ are small\cite{Acharya:2010af}
\be
c_1 \, \sim c_2 \, \sim \, \mathcal{O}(1/10 - 1/100)
\ee
are preferred. 
This would correspond, for example, to regions where the mediation scheme 
follows a mirage pattern\cite{mirage}.

\section{Summary and conclusions}
\label{sec:conclude}

In this paper we have implemented a statistical calculation of the SUSY breaking scale 
assuming a fertile patch of the string landscape where the low energy effective theory is comprised of the 
MSSM plus a hidden sector as described by $N=1$ $d=4$ supergravity with SUGRA assumed 
spontaneously broken via the super-Higgs mechanism. 
We have further assumed the existence of a vast array of scalar potential 
minima leading to different SUSY breaking scales. 
It is assumed that an assortment of SUSY breaking $F$ and $D$ terms are present and that their vevs 
are uniformly distributed. Such an assumption leads generally 
to the expectation of a landscape pull towards large values of hidden sector mass scales 
favored by a power law behavior $f_{SUSY}\sim (m_{hidden}^2)^{2n_F+n_D-1}$ which at first glance 
would seem to favor high scale SUSY breaking for $n=2n_F+n_D-1\ge 1$. If such were the case-- then provided
electroweak symmetry even breaks properly-- one would expect a value of the weak scale far beyond its
measured value characterized by $m_{W,Z,h}\sim 100$ GeV. 
A huge value of the weak scale would lead to far heavier particle masses and a suppression of 
weak interactions as we know them, and quite likely to a universe not conducive to 
complexity and life as we know it. 

As in Weinberg's estimate of the magnitude of the cosmological constant,
one may then assume an anthropic selection of weak scale values not-too-far from its measured value.
Requiring in addition a not-too-large value for the weak scale, corresponding to $\Delta_{\rm EW}\alt 30$ or
$m_{Z}\alt 350$ GeV (or $f_{EWFT}=\Theta (30-\Delta_{\rm EW})$), 
we are able to compute superparticle and Higgs mass probability distributions for
any assumed value of $n$. Remarkably, we find that for the simplest case, $n_F=1$, $n_D=0$ yielding a linear
draw of $f_{SUSY}\sim m_{soft}^1$, that the Higgs mass $m_h$ probability distribution is sharply peaked
at $m_h\simeq 125$ GeV. The $n=2$ result gives $m_h\sim 126$ GeV while a uniform scan corresponding 
to $n_F=0$ $n_D=1$ usually yields too low a value of $m_h$ (although $m_h\sim 125$ GeV 
is still possible-- see Fig. \ref{fig:3}). 
Values of $n\ge 3$ leads to a hard pull on soft terms that tend to place one in a situation 
with either CCB vacua or vacua without electroweak symmetry breaking. 
Thus, our results favor a rather simple hidden sector for SUSY breaking leading to $n\sim 1$ or 2. 
Higher $n\ge 3$ values as might be expected for instance from $F$-theory 
constructions\cite{Heckman:2010bq,Schafer-Nameki:2015bva} will have
difficulty in generating a proper breakdown of electroweak symmetry.

We also examined a different anthropic suppression factor 
$f_{SUSY}\sim \Delta_{\rm EW}^{-1}$ which penalizes large values of $m_{weak}$ but does not eliminate them.
This anthropic suppression allows for much higher SUSY breaking scales and typically
too large a value of $m_h$. A combination of the two-- 
$f_{EWFT}\sim \Theta (30-\Delta_{\rm EW})\cdot\Delta_{\rm EW}^{-1}$-- 
leads back to results similar to case {\bf A} with $m_h\sim 125$ GeV.

From our $n=1,\ 2$ results which favor a value $m_h\sim 125$ GeV, then we also expect
\begin{itemize}
\item $m_{\tg}\sim 4\pm 2$ TeV,
\item $m_{\tst_1}\sim 1.5\pm 0.5$ TeV,
\item $m_A\sim 3\pm 2$ TeV,
\item $\tan\beta \sim 13\pm 7$,
\item $m_{\tw_1,\tz_{1,2}}\sim 200\pm 100$ GeV and
\item $m_{\tz_2}-m_{\tz_1}\sim 7\pm 3$ GeV with
\item $m(\tq ,\tell )\sim 20\pm 10$ TeV (for first/second generation matter scalars).
\end{itemize}
These results can provide some guidance as to SUSY searches at future colliders and also a convincing rationale
for why SUSY has so far eluded discovery at LHC.  
They provide a rationale for why SUSY might contain its own decoupling solution to the
SUSY flavor and CP problems and the cosmological gravitino and moduli problems.
They predict that precision electroweak and Higgs coupling measurements should look very
SM-like until the emergence of superpartners at LHC and/or ILC.
They also help explain why no WIMP signal has been seen: dark matter may be 
higgsino-like-WIMP plus axion admixture
with far fewer WIMP targets than one might expect under a WIMP-only dark matter hypothesis.
The rather large value of $m_{3/2}$ expected from these 
results points perhaps towards mirage mediation\cite{mirage} as another lucrative scenario.

{\it Acknowledgements:} 
We thank Daniel Chung for a discussion.
This work was supported in part by the US Department of Energy, Office
of High Energy Physics. The computing for this project was performed at the OU Supercomputing Center
for Education \& Research (OSCER) at the University of Oklahoma (OU).


%
\end{document}